\documentclass[11pt, a4paper]{article}
\pdfoutput=1

\usepackage[utf8]{inputenc}
\usepackage{graphicx}
\usepackage{geometry}
\usepackage[english]{babel}

\usepackage{tikz}
\usetikzlibrary{decorations.pathmorphing}
\usepackage{pgfplots}
\usepgfplotslibrary{patchplots}
\usepackage{subcaption} 

\usepackage[numbers,sort&compress]{natbib}
\usepackage{epsfig}
\usepackage{amsmath}
\usepackage{amssymb}
\usepackage{amsfonts}
\usepackage{anysize}
\usepackage{latexsym}
\usepackage{xcolor}
\usepackage[pdftex,colorlinks=true,linkcolor=blue,citecolor=blue]{hyperref}
\hypersetup{linktocpage}
\usepackage{float}
\usepackage{dsfont}
\numberwithin{equation}{section}

\definecolor{ccqqqq}{rgb}{1,0.5,0}
\definecolor{uuuuuu}{rgb}{0.26666666666666666,0.26666666666666666,0.26666666666666666}
\definecolor{qqwwzz}{rgb}{0,0.3,0.9}

\newcommand{\beq}{\begin{equation}}
\newcommand{\eeq}{\end{equation}}
\newcommand{\bea}{\begin{eqnarray}}
\newcommand{\eea}{\end{eqnarray}}
\newcommand{\bit}{\begin{itemize}}
\newcommand{\eit}{\end{itemize}}
\def\nl{\nonumber \\}

\def\a{\alpha}

\def\b{\beta}

\def\p{\partial}

\def\le{\left(}
\def\ri{\right)}

\renewcommand{\a}{\alpha}
\renewcommand{\b}{\beta}

\date{}

\begin{document}

\begin{titlepage}

\begin{flushright}
  \hfill{YITP-25-31}
\end{flushright}

\bigskip
\begin{center}
{\LARGE  {\bf
Solitonic vortices and black holes  \\ with vortex hair in AdS$_3$ 
  \\[2mm] } }
\end{center}
\renewcommand{\thefootnote}{\fnsymbol{footnote}}
\bigskip
\begin{center}
 {\large \bf Roberto Auzzi$^{a,b}$},
 {\large \bf Stefano Bolognesi$^{c}$,}
   {\large \bf Giuseppe Nardelli$^{a,d}$}, \\
 {\large \bf Gianni Tallarita$^{e}$}  {\large \bf and} 
  {\large \bf Nicol\`o Zenoni$^{f}$} 
\vskip 0.20cm
\end{center}
\vskip 0.20cm 
\begin{center}
$^a${\it \small  Dipartimento di Matematica e Fisica,  Universit\`a Cattolica
del Sacro Cuore, \\
Via della Garzetta 48, 25133 Brescia, Italy}
\\ \vskip 0.20cm 
$^b${\it \small{INFN Sezione di Perugia,  Via A. Pascoli, 06123 Perugia, Italy}}
\\ \vskip 0.20cm 
$^c${\it \small{Department of Physics  E. Fermi, University of Pisa
and INFN Sezione di Pisa \\
Largo Pontecorvo, 3, Ed. C, 56127 Pisa, Italy}}
\\ \vskip 0.20cm 
$^d${\it \small{TIFPA - INFN, c/o Dipartimento di Fisica, Universit\`a di Trento, \\ 38123 Povo (TN), Italy} }
\\ \vskip 0.20cm 
$^e${\it \small{Departamento de Ciencias, Facultad de Artes Liberales,
Universidad Adolfo Ib\'a\~nez, \\  Santiago 7941169, Chile}}
\\ \vskip 0.20cm 
$^f${\it \small{Yukawa Institute for Theoretical Physics, Kyoto University, Kyoto 606-8502, Japan}}
\\ \vskip 0.20cm 
{\small E-mails: roberto.auzzi@unicatt.it, stefano.bolognesi@unipi.it, giuseppe.nardelli@unicatt.it, \\
gianni2k@gmail.com, nicolo@yukawa.kyoto-u.ac.jp}
\end{center}
\vspace{3mm}

\begin{abstract}

We study soliton and black hole solutions with scalar hair 
in AdS$_3$ in a theory with a Maxwell field 
and a charged scalar field with double trace boundary conditions,
which can trigger the dual boundary theory to become a superconductor.
We investigate the phase diagram as a function of the temperature $T$
and of the double trace coupling $\kappa$ and we find 
a rich pattern of phase transitions, which can be of the Hawking-Page
kind or can be due to the condensation of the order parameter.
We also find a transition between vortex solutions and the zero
temperature limit of the black hole for a critical value of
the double trace coupling $\kappa$.
The Little-Park periodicity is realized for the dual of the black hole
solution with hair  as a shift in the winding number and in 
the gauge field.

\end{abstract}

\end{titlepage}

\tableofcontents

\section{Introduction}

The AdS/CFT correspondence provides an interesting bridge between
gravitational systems  and superconductors \cite{Hartnoll:2008vx,Hartnoll:2008kx}.
The Coleman-Mermin-Wagner-Hohenberg (CMWH) theorem 
\cite{Coleman:1973ci,Mermin:1966fe,Hohenberg:1967zz}
prevents spontaneous symmetry breaking of a continuous symmetry
in  $2$-dimensional spacetime, and so it  prevents
the existence of superconductors in low-dimensional systems.
In holography, $1+1$ superconductivity  is possible due to the large $N$ limit \cite{Anninos:2010sq}.
In the case of the quantum field theory dual of a classical gravitational theory, the number of degrees of freedom per site
 is large and the fluctuations from mean field theory are parametrically suppressed.
Superconductivity in   systems with one space dimension has been 
studied both theoretically and experimentally in condensed matter
physics, see  \cite{review-superconductors-1-dim} for a review.
One-dimensional superconductors can be realized in nature 
by taking wires of superconducting material with thickness
 greater than the coherence length, which measures the size of the Cooper pairs. 
The mechanism that destroys  the long range order
of the superconductor  is  the fluctuations in the  phase of the order parameter.
Phase slips can be induced by a
suppression of the order parameter down to zero inside the wire,
which can be due both to thermal or quantum fluctuations \cite{phase-slips}.
Superconducting rings do not strictly contradict CMWH theorem, 
because superconductivity is always destroyed
for sufficiently small wire thickness.

The AdS/CFT correspondence provides a controlled theoretical laboratory
to study phase transitions in a strongly coupled regime.
The most famous example is the Hawking-Page transition \cite{Hawking:1982dh}
between global AdS and the spherical AdS black hole, which in the boundary field theory
can be interpreted as a thermal deconfinement phase transition \cite{Witten:1998zw}
for a Conformal Field Theory (CFT) on a sphere.
Another interesting example is the phase transition between AdS
soliton and AdS black brane \cite{Horowitz:1998ha,Surya:2001vj}
for a CFT on a cylinder. In this case, at low temperature the AdS
soliton has a lower free energy, due to the negative contribution to the Casimir energy
which comes from the anti-periodic boundary conditions for the fermions.
For the  AdS$_3$ spacetime, the AdS soliton coincides with the global 
AdS background and corresponds to a CFT on a circle $S^1$ at finite temperature,
  and so the two transitions are identical.

In the framework  of holographic superconductors in AdS$_4$, 
the phase diagram of the dual CFT on a cylinder has a richer phase structure
 \cite{Nishioka:2009zj,Horowitz:2010jq} as a function of temperature and chemical potential.
Here the chemical potential is the external parameter used to induce the condensation of the scalar field. 
Apart from the  global AdS and the  AdS black hole, 
two other phases can be realized, corresponding to the  AdS
soliton, with condensate and no horizon, and black hole dressed with scalar hair, respectively.
This is a simple setup in which one can realize 
a complex competition among several quantum ground states.
We will study an analogous phase diagram
in a model of AdS$_3$ holographic superconductors, as 
a function of the temperature and of a double trace coupling $\kappa$, which is 
the parameter that  triggers the condensation of the scalar order parameter in our setup \cite{Witten:2001ua,Berkooz:2002ug,Faulkner:2010gj}.

Another class of transitions involves  solitons and  black holes.
In particular, in four dimensional flat spacetime the phase transition 
between monopole solutions and black holes was  investigated in \cite{Lee:1991vy,Ortiz:1991eu,Breitenlohner:1991aa,Breitenlohner:1994di,Lue:1999zp,Weinberg:2001gc}.
In this case, a  monopole soliton exists only if the vacuum expectation value
of the scalar field $v$ is less than a critical value $v_{cr}$.
Nearby $v_{cr}$, the monopole solution approaches a Reissner-Nordstr\"om
extremal black hole. We will investigate a similar transition
in AdS$_3$ between  vortex solitons and the zero temperature limit
of a family of hairy black hole solutions.

In this paper, we will investigate various phase transitions 
in three dimensional AdS gravity coupled to a Maxwell
and to a charged scalar field. 
We first discuss configurations with vanishing scalar field,
which are the analog of the 
 Melvin flux tube in flat space \cite{Melvin:1963qx}.
 In this case, an analytical solitonic solution which includes gravitational 
 backreaction  \cite{Clement:1993kc,Hirschmann:1995he,Cataldo:1996ue,Dias:2002ps}
 can be found by analytic continuation of the charged BTZ black hole  
 \cite{Banados:1992wn,Banados:1992gq,Martinez:1999qi}.
We will discuss the phase diagram of the system,
as a function of the external source for the gauge field.

We further address a class of  black hole 
solutions with vortex hair  in  asymptotically AdS$_3$ spacetime.
In the framework of holographic superconductors, the 
boundary dual of a black hole solution with vortex hair
\cite{Montull:2011im,Montull:2012fy} realizes  the Little-Parks (LP) periodicity \cite{Little-Parks},
 which tells that the current  flowing in a superconducting ring is periodic as a function of the
magnetic flux measured in the perpendicular direction to the plane of the ring.
For our black hole solution, the LP periodicity is realized as a symmetry
which shifts the winding number $n$ of the vortex and the expectation value
of the Wilson line. We further investigate the zero temperature limit of the hairy black hole solution,
which is a configuration with zero entropy and a mild logarithmic singularity.
This solution is similar to the one studied in AdS$_4$ in \cite{Horowitz:2009ij}.

We also study soliton solutions, which are an AdS version of
the Abrikosov-Nielsen-Olesen (ANO) vortices \cite{Abrikosov:1956sx,Nielsen:1973cs}. 
The solution with vanishing winding number $n=0$
corresponds to the vacuum.
In order for the topological soliton with winding $n$
to exist, we need $\kappa<\kappa_{c1}$, where
$\kappa_{c1}$ is a negative threshold coupling, which is function of the winding $n$.
For vortex solutions with $n \geq 1$, we find that there is
another critical value $\kappa_{c2}$ of the double trace coupling  
below which the soliton solution no longer exists.
Approaching the critical value $\kappa \to \kappa_{c2}$, there is a transition
between the vortex and the zero temperature limit 
of the hairy black hole solutions, which is similar to
the one between magnetic monopoles and extremal
Reissner-Nordstr\"om black holes in flat space \cite{Lee:1991vy,Ortiz:1991eu,Breitenlohner:1991aa,Breitenlohner:1994di,Lue:1999zp,Weinberg:2001gc}.
A vortex soliton with non-vanishing winding $n$ exists
in a limited window of the coupling $\kappa_{c2} < \kappa < \kappa_{c1}$,
which is a function of $n$.
Contrary to what happens in flat space for the ordinary ANO vortex, the magnetic flux of a vortex soliton in 
general is not quantized. 
The shift symmetry which realizes the LP periodicity
 in the black hole phase is explicitly broken by the 
 boundary conditions for the gauge field at the center of the vortex soliton.

The paper is organized as follows. In Sec. \ref{sec-model} we introduce the model
and the ansatz for the solutions and we find an expression for the energy density
in the boundary theory, using holographic renormalization.
 In Sec. \ref{section-senza-scalare} we review a class of solutions 
   \cite{Clement:1993kc,Hirschmann:1995he,Cataldo:1996ue,Dias:2002ps}
with vanishing scalar field  and we discuss their phase diagram.
  In Sec. \ref{sect-BH-vortex-hair} we consider black hole solutions with 
  scalar vortex hair.   In Sec. \ref{sec-soliton} we consider a class of solutions 
  with scalar hair which are  solitonic vortices in AdS$_3$, which do not have
  a horizon and are regular everywhere. 
   We conclude in Sec. \ref{sec-conclusion}. 
   In the Appendices we provide some technical material.


\section{Theoretical setting}
\label{sec-model}

In this section we introduce the holographic model and 
an ansatz for a vortex-like configuration, which can be used to describe
both a solitonic solution or a black hole. 
We discuss the boundary conditions
for the scalar and the vector field and we use holographic
renormalization to find the relation between the asymptotic 
expansion of the profile function and the energy of the state
in the dual quantum field theory. 

Topological solitons in AdS have been studied by many authors.
Monopoles  in AdS space have been discussed by several authors, 
see for example  \cite{Lugo:1999fm,Lugo:1999ai,Bolognesi:2010nb,Esposito:2017qpj,Zenoni:2021iiv}.
AdS vortex solitons were  recently studied
in Refs. \cite{Edery:2020kof,Albert:2021kne,Edery:2022crs},
 in a setting where the scalar  breaking the $U(1)$ 
 symmetry acquires a bulk expectation value which is not vanishing at the boundary.
  This is a case of explicit symmetry breaking in the dual theory, which is permitted
   if $U(1)$ is a global symmetry in the dual theory. 
   A black hole solution with vortex hair was found in  \cite{Cadoni:2009cq},
in a setting with no gauge field.
Examples of hairy black hole solutions with an electric charge in asymptotically AdS$_3$ spacetime
were studied in \cite{Priyadarshinee:2023cmi,Daripa:2024ksg}.

\subsection{The model}
We consider Einstein gravity coupled to a complex scalar field $\phi$ 
and a $U(1)$ gauge field $A_{\mu}$ in $3$ dimensions with negative cosmological constant
\beq
\begin{aligned}
& S = \int  \left[ \frac{1}{16 \pi G} \le R - 2 \Lambda \ri + \mathcal{L}_{\rm{m}} \right]  \sqrt{-g} \, d^3x
+\frac{1}{8 \pi G} \int_{\p} K   \sqrt{-h} \, d^2x
\, , \\
& \mathcal{L}_{\rm{m}} = -\frac{1}{4} F^{\mu \nu} F_{\mu \nu} 
 - \le D^{\mu} \Phi \ri^* D_{\mu} \Phi 
 - m^2 \, \Phi^* \Phi   \, ,
\end{aligned}
\label{modello}
\eeq
where
\beq
D_{\mu} \Phi  =\p_{\mu}  \Phi- i e A_{\mu} \Phi \, ,
\qquad
F_{\mu \nu}  = \p_{\mu} A_{\nu} - \p_{\nu} A_{\mu}
\eeq
and $e$ is the gauge coupling constant. 
The boundary part in eq.~(\ref{modello}) is the Gibbons-Hawking term,
where $K$ the extrinsic curvature
and  $h$ the determinant of the induced metric at the boundary.
This theory  was studied  as a model for 
two-dimensional holographic superconductors, see for example
  \cite{Hung:2009qk,Ren:2010ha,Sonner:2014tca}.

The equations of motion for the matter are
\beq
\label{eom-maxwell}
 D^{\mu} D_{\mu} \Phi - m^2 \, \Phi = 0 \, , 
\qquad
 \nabla^{\mu} F_{\mu \nu} +J_\nu  = 0 \, ,
\eeq
where $J_\nu$ is the bulk electric current
\beq
J_\nu=i e \Phi \le D_{\nu} \Phi \ri^* - i e \Phi^* D_{\nu} \Phi \, .
\eeq
In eq.~(\ref{eom-maxwell}) we denote by  $\nabla_{\mu} $
the gravitational covariant derivative and by
 $D_{\mu} $  a combination of  both gravitational and gauge covariant derivatives.
The Einstein equation is
\beq
\label{eom-einstein}
 R_{\mu \nu} - \frac{1}{2} g_{\mu \nu} R + \Lambda g_{\mu \nu} = 8 \pi G \, T_{\mu \nu} \, ,
\eeq 
where $T_{\mu \nu}$ the bulk energy-momentum tensor is
\beq
\label{bulk_stress}
T_{\mu \nu} = g^{\rho \sigma} F_{\mu \rho} F_{\nu \sigma} 
+ \le D_{\mu} \Phi \ri^* D_{\nu} \Phi + D_{\mu} \Phi \le D_{\nu} \Phi \ri^* + g_{\mu \nu} \, \mathcal{L}_{\rm{m}} \, . 
\eeq
We denote the coordinates as
\beq
x^\mu=(r,t,\varphi) \, ,
\eeq
where $0 \leq \varphi \leq 2 \pi$.
The  solution with lowest energy in the vacuum $T_{\mu \nu} = 0$ is  global AdS$_3$
\beq
ds^2 = L^2 \left[ -\left(1+{r^2}\right)  d t^2 +   \frac{dr^2}{1+{r^2}} + r^2 \, d\varphi^2   \right] \,
\eeq 
where the  curvature radius $L$  is related to the cosmological constant as follows
\beq
\Lambda = -\frac{1}{L^2} \, .
\eeq
The central charge of the dual CFT is
\beq
c=\frac{3 L}{2 G} \, .
\label{central-charge}
\eeq
We set the AdS radius $L=1$ from now on.

\subsection{Ansatz and equations of motion}

The vortex solution is given by a generalisation of the Abrikosov-Nielsen-Olesen vortex 
\cite{Abrikosov:1956sx,Nielsen:1973cs} ansatz
\beq
\label{NO_ansatz}
\Phi =  H(r) \, e^{i n \varphi} \, , 
\qquad
 A_\mu dx^\mu \, =   \frac{a(r)}{e}  \, d\varphi \, ,
\eeq
where the integer number $n $ describes the 
winding  of the scalar field. For $n=0$, the field configuration
 in eq.~(\ref{NO_ansatz})   describes the vacuum of the theory.
  The configurations with $n>0$ instead correspond to vortices.
We take the following spherically symmetric ansatz for the metric:
\beq
\label{metric_generica}
ds^2 = -(1+r^2) h(r) g(r) d t^2 + \frac{h(r)}{g(r)} \frac{dr^2}{1+r^2} + r^2 \, d\varphi^2  \, .
\eeq 
We  require 
\beq
\lim_{r \rightarrow +\infty} h(r) = \lim_{r \rightarrow +\infty} g(r) = 1
\eeq
which fixes the asymptotic normalization of the  coordinate $t$.
Empty global AdS$_3$ is recovered for $h=g=1$.
It is convenient to introduce the auxiliary function
\beq
q(r)=(1+r^2) \, g(r) \, .
\eeq
Inserting the ansatz  (\ref{NO_ansatz})
and (\ref{metric_generica}) in eqs.~(\ref{eom-maxwell}) and (\ref{eom-einstein})
we find the equations of motion 
\beq
\begin{aligned}
& \frac{d}{dr} \le q \, r  H' \ri = h \,  H \le \frac{ (n-a)^2}{r} + r \, m^2  \ri , \\
& \frac{d}{dr} \le \frac{q }{r} \, a' \ri = 2 e^2 \frac{a-n}{r} \, h\,H^2 \, , \\
& \frac{h'}{h} = 8 \pi G \, \le 2 r {H'}^2 + \frac{ {a'}^2}{e^2  \, r} \ri  \, , \\
& q' = h \le 2 r - {16 \pi G} \le \frac{(a-n)^2}{r} + m^2  r \ri  \, H^2 \ri \, . 
\end{aligned}
\label{eom-generiche-2}
\eeq
It is also convenient to introduce  $\rho$, which
 is the matter energy density for the soliton solution
and outside the horizon of the black hole solution
 \beq
 \rho=T_{\mu \nu} n^{\mu} n^{\nu} \, ,
 \eeq
where  $n^\mu$ is the unit normal
to the constant $t$ hypersurfaces.
Note that $\rho$ is not the matter energy density 
inside  the horizon of the black hole solution, because there
$n^\mu$ is not timelike.
A direct evaluation gives
 \beq
   \rho = -\mathcal{L}_{\rm m} = \frac{g}{h} \,  (1+r^2) \, \le {H'}^2+
\frac{1}{2 e^2} \frac{ {a'}^2 }{r^2}  \ri
 +  H^2 \left( \frac{(n-a)^2}{r^2} +m^2 \right) \, .
 \label{densita-energia-generica}
 \eeq 
The matter part of the action computed on the
 ansatz in eqs.~(\ref{NO_ansatz}) and (\ref{metric_generica}) is
  \beq
  S_{\rm m}=   \int \mathcal{L}_{\rm{m}} \, \sqrt{-g} \, d^3 x=-2 \pi \, \int \rho \, h \, r \, dr  \, dt \, .
  \label{azione}
  \eeq
From the constraint equations (see \cite{Poisson-book} for a review)
we  find the following relation  between the
 the Ricci curvature of constant-time hypersurfaces $^2 R $
 and the energy density $\rho$ 
\beq
\,^2R = -2 +16 \pi G \, \rho \, \, .
\label{curva-2dim}
\eeq
In order for eq.~(\ref{curva-2dim}) to be valid,
 we need the property that the extrinsic curvature 
tensor vanishes on a constant-$t$ hypersurface,
which holds because the spacetime is static.
 The 2-dimensional Ricci scalar is
\beq
^2 R \, r= - \frac{d}{dr} \le  \frac{q }{h} \ri \, .
\label{curva-2dim-bis}
\eeq
Comparing eq.~(\ref{curva-2dim}) and  (\ref{curva-2dim-bis})  we find the useful relation
\beq
 \frac{d}{dr} \le   \frac{q}{h} \ri  = (2 -16 \pi G \, \rho) \,  r \, ,
 \label{constraint-specializzata}
\eeq
which indeed can be proved directly using the equations of motion
eqs.~(\ref{eom-generiche-2}).

We will consider two classes of solutions:
\begin{itemize}
\item
Soliton solutions, which are smooth everywhere in the bulk.
To avoid a singularity for the $U(1)$ gauge field at $r=0$,
 we require that $a(r)$ vanishes quadratically as 
 \beq
 a(r) \sim r^2 \, .
 \label{no-string}
 \eeq
At $r = 0$, we have also  to impose 
\beq 
h(0) = g(0)
\label{no-cone}
\eeq
 in order to avoid a conical singularity in the geometry. 
For $r \to0$, we can approximate the equation for $H$ 
in eq.~(\ref{eom-generiche-2}) as follows 
\beq
H''=-\frac{H'}{r}+n^2 \, \frac{H}{r^2} \, ,
\eeq
which has as solution 
\beq
H(r) \approx H_0 \, r^n \, .
\label{H-origine}
\eeq
For $n=0$, we have that the maximum of the scalar field $H$
is at $r=0$, see the numerical solutions in figure \ref{plot-H-a-hh-g} for a plot. 
For higher windings $|n| \geq 1$, instead, $H$ vanishes at $r=0$.
 
\item
Black hole solutions instead
 do not need to be smooth for $r \to 0$. 
In this case, we denote by $r_h$ the radius of the
event horizon, for which
\beq
g(r_h)=0 \, .
\eeq
\end{itemize}
In order to find numerical solutions to the equations of motion,
it is useful to introduce the compact coordinate 
\beq
\psi = \arctan r \, .
\label{coordinata-psi}
\eeq
See appendix~\ref{appe-psi} for the equations of motion
in the $\psi$ coordinate.

\subsection{Boundary conditions for the scalar}

For generic mass $m$, we have that the dimensions of the dual operators are
\beq
\Delta_\pm=1 \pm  \sqrt{1+  \, m^2 } \, 
\eeq 
and the  Breitenlohner-Freedman (BF) bound \cite{Breitenlohner:1982jf} 
gives $m^2 \geq -1$. We will later focus
 on the mass window
\beq
-1 < m^2  < -\frac{3}{4 } \, ,
\label{massa34}
\eeq
in order to keep as simple  as possible
the technical details of holographic renormalization, see section~\ref{appe-boundary-stress-tensor}.
The scalar field at the boundary can be expanded as
\beq
\Phi(r) = \frac{\phi^0}{r^{\Delta_-}} + \frac{\phi^1}{r^{\Delta_+}} + \dots \, ,
 \label{asintotica-campo-scalare}
\eeq
or, in terms of $H(r)$
\beq
H(r) = \frac{\alpha_H}{r^{\Delta_-}} + \frac{\beta_H}{r^{\Delta_+}} + \dots \, .
 \label{asintotica-massa-generica}
\eeq
Since the mass of the scalar field falls into the range $-1 \leq m^2  \leq 0$,
 both modes in the scalar field expansion are normalizable. 
In the standard (Dirichlet) quantization, $\a_H$ is the source
for the dual operator $\mathcal{O}_+$ which has dimension $\Delta=\Delta_+$,
and $\b_H$ is proportional to the VEV of the operator.
In the alternative (Neumann) quantization, where $\b_H$ is the source
and $\a_H$ is proportional to the VEV, we have that the 
dual operator $\mathcal{O}_-$ has dimension $\Delta=\Delta_-$.

  In order to consider a spontaneous $U(1)$ symmetry breaking,
 we   work with double trace deformations, as in 
 \cite{Witten:2001ua,Berkooz:2002ug,Faulkner:2010gj,Dias:2013bwa}. 
 In this way we can realize  spontaneous symmetry breaking,
 triggering the condensation of the expectation value of 
 an operator of the dual theory.
 This correspond to working with the alternative quantization,
with an extra term in the hamiltonian density
\beq
\Delta \mathcal{H} =\kappa \, \mathcal{O}_- \mathcal{O}_-^\dagger \, .
\label{H-double-trace}
\eeq
A negative $\kappa$ tends to drive the system to the condensation
of the order parameter.
The coupling  $\kappa$ in the dual conformal field theory has dimension
 \beq
 [\kappa]= 2 \, \sqrt{1+m^2 }  \, .
 \label{dimensione-kappa}
 \eeq
The  double trace conditions are implemented as the following
 boundary condition in the gravity dual
  \beq
 \phi^1=\kappa \, \phi^0 \, ,
 \qquad
 \b_H= \kappa \,  \a_H \, ,
 \eeq
 and the vacuum expectation value
 of the operator $\mathcal{O}_+$  is proportional to $\phi^0$.
There is a critical negative value of $\kappa$
 for which the solution starts developing hair.

 Solutions with scalar hair which are similar to the ones discussed in this paper
 have been previously studied in higher dimensional AdS spacetime
 \cite{Nishioka:2009zj,Horowitz:2010jq,Montull:2011im,Montull:2012fy},
 in a setup where the condensation of order parameter is due to
 chemical potential and non-zero charge density.
 In this case, the mechanism that drives the condensation of the scalar field
 is a negative contribution to the effective mass squared of the scalar,
due to the $-e^2 \, g^{tt} (A_t)^2 |\Phi|^2 $ term in the Lagrangian of a charged scalar field.
This mechanism requires the presence of electric charge in the bulk. 
We instead focus on an alternative mechanism to induce the condensation of the 
scalar field, which is to add a double trace deformation, as proposed in \cite{Faulkner:2010gj}.
This mechanism does not require the presence of 
 a background charge density and it works also for Schwarzschild AdS.
 Also, it does not break the Lorentz symmetry on the boundary.
The double trace coupling $\kappa$ is identified with the coefficient of the square of
 the order parameter, as in the weakly coupled Landau-Ginzburg model.
 If the microscopic description of the operator $\mathcal{O}$ is the trace of a fermion bilinear,
 $\kappa$ corresponds to a four-fermion interaction.

\subsection{Boundary conditions for the gauge field}

Boundary conditions for vector fields play an important role in AdS/CFT
correspondence \cite{Witten:2003ya,Marolf:2006nd}. 
In AdS$_d$ with $d\geq 4$ with Dirichlet boundary conditions,
the bulk $U(1)$ gauge invariance corresponds to a global $U(1)$
symmetry in the boundary CFT. For $d=4$, Neumann boundary conditions
for the Maxwell vector field are dual to a gauged $U(1)$ in the boundary field 
theory. In AdS$_4$, we can use the Maxwell theory with a charged scalar 
to describe both superfluids or superconductors, depending on the 
choice of Dirichlet or Neumann boundary conditions for the $U(1)$
gauge field \cite{Domenech:2010nf}. 

In AdS$_3$, the bulk dual of a global $U(1)$
symmetry on the boundary field theory
 is provided by a vector field with a Chern-Simons term \cite{Jensen:2010em}.
The Maxwell vector field in AdS$_3$ instead is a bit special.
Let us work in the gauge  where $A_r=0$.  The asymptotic expansion
of the $U(1)$ gauge field is
\beq
A_\mu= \tilde{\mathfrak{a}}_\mu \log \frac{r}{r_a} + \mathfrak{a}_\mu + O\Big(\frac{1}{r}\Big) \, ,
\eeq
where $r_a$ can be interpreted as a renormalization scale.
The field $A_\mu$ is dual to a gauged symmetry in the boundary theory
\cite{Jensen:2010em,Faulkner:2012gt},
where $\tilde{\mathfrak{a}}_\mu$ plays the role of an external conserved current
\beq
\p_\mu \tilde{\mathfrak{a}}^\mu=0 \, ,
\label{conservazione-source}
\eeq
 and  $\mathfrak{a}_\nu$ is a dynamical gauge field coupling to this current.
 
 If $ \tilde{\mathfrak{a}}^\mu$ is interpreted as a generic source, which does not satisfy the conservation 
 law in eq.~(\ref{conservazione-source}), there are some subtleties with holographic
 renormalization   \cite{Argurio:2016xih}. In particular, a counterterm
 which scales as  ${1}/{\a_H}$  (which is diverging with Dirichlet 
 boundary conditions for the scalar field)
 is needed, see 
  Ref. \cite{Argurio:2016xih}.
 In this paper, we will always enforce the conservation law eq.~(\ref{conservazione-source})
 for the external current. Also, we will consider double trace boundary
 conditions for the scalar, and so $\a_H$ does not vanish in backgrounds
 with non-vanishing scalar field.

In three dimensions, a gauge field $A_\mu$ can be dualized to a massless 
scalar.  In the absence of a charged scalar field
 we can rewrite  (as proposed in  \cite{Marolf:2006nd,Jensen:2010em,Faulkner:2012gt})
  the Maxwell action in terms of an auxiliary real scalar $\theta$
 \beq
 \p^\mu \theta=  \frac{\epsilon^{\mu \a \b}}{\sqrt{-g}} \p_\a A_\b 
\, , \qquad
  F^{\mu \nu} =\frac{\epsilon^{\mu \nu \rho}}{\sqrt{-g}} \p_\rho \theta \, .
  \label{scalar-duality}
 \eeq
 The asymptotic expansion of $\theta$ nearby the boundary is
 \beq
 \theta=\theta^{(0)} +\frac{\theta^{(2)} +\frac12 \Box^{(0)}  \theta^{(0)} \log r}{r^2} \, ,
 \eeq
 where $\Box^{(0)}$ is the boundary D'Alembert operator.
 We can  identify
\beq
\tilde{\mathfrak{a}}^\mu = \epsilon^{\mu \nu} \, \p_\nu \theta^{(0)} \, ,
\qquad
{f}^{\mu \nu} = \p^\mu \mathfrak{a}^\nu - \p^\nu \mathfrak{a}^\mu =-2 \, \epsilon^{\mu \nu} \, \theta^{(2)} \, .
\eeq
Taking $\tilde{\mathfrak{a}}_\mu $ as a source corresponds to the standard quantization
 for the dual scalar with source $\theta^{(0)}$.
 In the case studied in this paper, a charged scalar field is also present
 and the change of variables in eq. (\ref{scalar-duality})
 gives rise to a non-local action. 
Still, we will  use the duality as a consistency check
 for the holographic renormalization counterterms.

In our vortex ansatz, we can expand the function $a(r)$ nearby the boundary as follows
\beq
 \qquad a(r) = \tilde{a} \log \frac{r}{r_a} + a_0 + \dots \, .
 \label{asintotica-a}
\eeq
For $\tilde{a} \neq 0$, there is a renormalization ambiguity in $a_0$
due the presence of the scale $r_a$.  This ambiguity disappears for $\tilde{a}=0$.
In the following, for simplicity we will often set $r_a=1$.

It is useful to consider the bulk magnetic field of the vortex
\beq
F= \frac{a'(r) }{e} \, dr \wedge d \varphi \, .
\eeq
The flux of the magnetic field is given by
\beq
\mathcal{F}=\frac12 \int F_{\mu \nu} d x^\mu \, dx^\nu = \oint A_\mu dx^\mu = \frac{2 \pi }{e} a(\infty) \, .
\label{formula-flusso}
\eeq
For a vortex in flat space, we have that the flux is quantized 
because we need  $D_\mu \phi \to 0$ in order for the energy to be finite.
The flux instead is not quantized for the AdS vortex solution that
we consider in this paper. For non-zero $\tilde{a}$, we have that the flux
 is infinite. For $\tilde{a}=0$ the flux is 
 proportional  to $a_0$  and it is finite, but not necessarily quantized.
In our background, we have that the term in the lagrangian 
density in eq.~(\ref{densita-energia-generica})
which is proportional to $(n-a)^2$ is
\beq
H^2 \le\frac{n-a}{r}\ri^2 \,  \approx \, 
 \a_H^2 \frac{(n-a)^2}{r^{2(2-\sqrt{1+m^2})}}  \, ,
 \label{stima-energetica}
\eeq
where $\approx$ is the leading behaviour at large $r$.
This term does not give rise to a divergence when integrated in the action eq.~(\ref{azione})
  for $m^2 < 0$, which is always realized in the mass range studied in this paper, see eq.~(\ref{massa34}).

For soliton solutions, our background has always vanishing electric field.
For black hole solutions, $F_{r \varphi}$ is still a magnetic field for $r>r_h$,
but instead it becomes an electric field for $r<r_h$,  because inside the horizon
 $r$ becomes a timelike coordinate.
Since for our solution we have that at a given point either the electric or the magnetic
field vanishes, it is convenient to use the quantity
\beq
B^2= \frac{e^2}{2} \, F_{\mu \nu} F^{\mu \nu} 
=\frac{(1+r^2) g}{h}   \, \le \frac {a'}{r} \ri^2
=\frac{g}{h}  \,  \frac{\cos^4 \psi}{\sin^2 \psi}  \, (a'(\psi))^2
\label{Bq-espressione}
\eeq
to measure the strength of the magnetic or electric field at a given point.
If $B^2$ is positive, it is proportional to the square of the magnetic field.
In a region where $B^2$ is negative, it is proportional to minus the square of the
electric field.

\subsection{Holographic renormalization}
\label{appe-boundary-stress-tensor}

In order to compute the energy of soliton and black hole solution,
we use holographic renormalization, see \cite{Balasubramanian:1999re,Skenderis:2002wp,Henningson:1998gx}.
We work in the gauge where $A_r=0$.
The expansion of the metric profile functions nearby the boundary, 
in the mass window in eq.~(\ref{massa34}), is
\bea
&& h(r) = 1 - \frac{ g_1 }{r^{2 \Delta_-}} 
+\frac{ h_2 }{r^2} 
+ \dots \, ,
 \nl
&& g(r) = 1+ \frac{g_1}{r^{2 \Delta_-}} 
+\tilde{g}_2 \, \frac{1}{r^2} \log \frac{r}{r_g}
+ \frac{g_2}{r^2} + \dots \, ,
\label{espansione-boundary}
\eea
where $r_g$ is a radial scale which is similar to the scale $r_a$
in eq.~(\ref{asintotica-a}).
For $\tilde{g}_2 \neq 0 $ there is an ambiguity in $g_2$ due to the presence 
of the scale $r_g$.
Outside the window in eq.~(\ref{massa34}),
for $m^2\geq -\frac{3}{4}$, the expansion in eq.~(\ref{espansione-boundary})
becomes more complicated, and it will not be discussed here.
Inserting the asymptotic expansions in eqs.~(\ref{espansione-boundary}),
(\ref{asintotica-massa-generica}) and (\ref{asintotica-a}) in the equation of motion
eq.~(\ref{eom-generiche-2}), we find the following relations
\beq
g_1=8 \pi G \, \Delta_- \, \a_H^2 \, ,
\qquad
\tilde{g}_2=-8 \pi G \, \frac{  \tilde{a}^2}{e^2}\, ,
\qquad
h_2= 16 \pi G \, m^2 \, \a_H \b_H 
-4 \pi G \, \frac{ \tilde{a}^2}{e^2} \, .
\eeq
Note that  $g_2$ is not fixed by the asymptotic expansion of the equations of motion.

It is useful to introduce the inward pointing unit normal to the boundary
\beq
n^{\mu} = (n^r,0,0) \, , \qquad 
n^r=-\sqrt{\frac{1}{g_{rr}}} \, .
\eeq
We denote by $\gamma_{ab}$  the metric induced on  constant $r$
hypersurfaces near the boundary
\beq
\gamma_{ab}= \left(
\begin{array}{cc}
g_{tt} & 0 \\
0 & r^2 \\   
\end{array} 
\right) \, .
\eeq
The counterterms  in the action \cite{Balasubramanian:1999re} are  
\beq
 S_{ct}= \int_{r=r_\Lambda}  d^2 x \sqrt{-\gamma} \left( -\frac{1}{8 \pi G} - \Delta_- \Phi^* \Phi 
 -n_\a A_\mu F^{\a \mu}
 -\frac{F_{r a} F^{r a}}{2} \, \log {\frac{r}{r_R}} \right) \, ,
 \label{counterterms}
\eeq
where $r_R$ is a renormalization scale,
 $\gamma$ is the determinant of the 
 induced metric at the boundary $\gamma_{ab}$
 and $r_\Lambda$ is the UV cutoff.
The fourth term in eq. (\ref{counterterms}) was introduced  in  \cite{Jensen:2010em}
in order to cancel the UV divergencies due
to the $U(1)$ gauge field.
This term reproduces  the scalar counterterm 
in the dual formulation in eq. (\ref{scalar-duality}),
which is
\beq
  S_{\theta,ct}=\frac{1}{2} \int_{r=r_\Lambda} d^2 x \sqrt{-\gamma} \left(
 \gamma^{ab} \p_a \theta \,  \p_b \theta  \right) \, \log \frac{r}{r_R} \, .
\eeq
Moreover, the fourth term in eq. (\ref{counterterms}) is equivalent 
(in the gauge where $A_r=0$) to the alternative counterterm
\beq
  S_{\theta,alt}=\frac{1}{2} \int_{r=r_\Lambda} d^2 x \sqrt{-\gamma}   \frac{A_\mu A^\mu}{\log (r/r_R) } \, ,
\eeq
which was used in \cite{Hung:2009qk,Ren:2010ha}.

The energy-momentum tensor with Dirichlet boundary condition  is
\bea
&&T^D_{ab} =  \frac{ K_{ab}  }{8 \pi G} 
-\gamma_{ab} \le \frac{K+1}{8 \pi G}+ \Delta_- \Phi^* \Phi
+n_\a A_\mu F^{\a \mu}
+ \frac{F_{r c} F^{r c}}{2} \, \log \frac{r}{r_R}  \ri
\nl
&& \qquad \quad 
- 2 \, n_\a \, A_{(a} F^{\a}_{\,\, \,\,\, b)}
-F_{r a} F^{r}_{\,\,\, b} \, \log \frac{r}{r_R}
\, ,
\eea
where $K_{ab}= n^{\mu} \partial_{\mu} \gamma_{ab}/2$ 
  is the extrinsic curvature tensor and $K= \gamma^{ab} K_{ab}$ is its trace. 

The Dirichlet energy density $\mathcal{E}^D$ and the pressure $p^D$ are
\bea
\label{T-dirichlet-massa-minore-3quarti}
&&\mathcal{E}^D=T^D_{tt}=-\frac{1+g_2}{16 \pi  G} 
 +(m^2+ 2 \Delta_-)\alpha _H  \beta_H -\frac{ \tilde{a}^2}{4 e^2} 
 -\frac{\tilde{a} \, a_0}{e^2} +\zeta \, ,
\nl
&&p^D= T^D_{\varphi \varphi }=-\frac{1+g_2}{16 \pi  G}
 -(3 m^2 +2 \Delta_-) \alpha _H  \beta_H + \frac{  \tilde{a}^2}{4 e^2}
 -\frac{\tilde{a} \, a_0}{e^2}
+\zeta  \, ,
\eea
where
 \beq
\zeta= \frac{\tilde{a}^2}{e^2} \log \frac{r_a}{\sqrt{r_g r_R}} \, .
\label{zeta}
 \eeq
The dependence of the energy momentum
 tensor on the renormalization scale is all in terms of $\zeta$.
In particular, energy differences measured at constant source $\tilde{a}$
are independent of the scales $r_{a,g,R}$. In the following, for simplicity we will set
\beq
r_a=r_g=r_R=1
\eeq
and then  $\zeta=0$.
For the case of empty global AdS, eq.~(\ref{T-dirichlet-massa-minore-3quarti})
reproduces the Casimir energy density $-\frac{1}{16 \pi G}$ for anti-periodic  fermionic
boundary conditions \cite{Horowitz:1998ha}.
Note that the trace anomaly $T=T^k_k$ is  
independent from the regularization parameter $\zeta$.

In this paper we will impose double trace boundary conditions, 
see eq.~(\ref{H-double-trace}).
  The double trace deformation induces  the following term 
  in the action of the boundary theory
 \beq
 \mathcal{F}= -\kappa \,\bar{\phi}^0  \phi^0 \, ,
 \eeq
where $\phi^0$ and $\phi^1$ refer to the
 asymptotic expansion in eq.~(\ref{asintotica-campo-scalare}).
 We take as external source
\beq
J_{\mathcal{F}}=-\phi^1- \p_{\bar{\phi}^0}\mathcal{F}=-\phi^1+ \kappa \, \phi^0
\eeq
 and add the following term to the renormalized action
 \beq
 S_{\mathcal{F}}=
 \int  d^2 x \sqrt{-h} \left( \bar{J}_{\mathcal{F}} \phi^0 + \mathcal{F} (\phi^0 )\right) + {\rm c.c. }\, .
 \eeq
 In this case, $\phi^0$ should be interpreted as VEV.
 Due to the shift in the action, we have that the energy tensor for 
 double trace boundary conditions (see  \cite{Papadimitriou:2007sj,Caldarelli:2016nni}) is
  \bea
T_{ij}&=& T_{ij}^{(D)}+\eta_{ij} [\bar{J}_{\mathcal{F}} \phi^0 + \mathcal{F} (\phi^0 ) +{\rm c.c. }]
\nl
 &=&
T_{ij}^{(D)}-\eta_{ij} [ 2 \alpha_H \beta_H ] \, .
\label{double-trace-mass-shift}
\eea
Combining eq.~(\ref{T-dirichlet-massa-minore-3quarti})
with eq.~(\ref{double-trace-mass-shift}), we find
energy and pressure for double trace boundary conditions
\bea
\label{T-double-trace-massa-minore-3quarti}
&& \mathcal{E} =-\frac{1+g_2}{16 \pi  G} 
 +(m^2+ 2 \Delta_- + 2) \alpha _H  \beta_H -\frac{ \tilde{a}^2}{4 e^2}
  -\frac{\tilde{a} \, a_0}{e^2} \, ,
\nl
&& p = -\frac{1+g_2}{16 \pi  G}
 -(3 m^2 +2 \Delta_- +2) \alpha _H  \beta_H
 + \frac{ \tilde{a}^2}{4 e^2}
  -\frac{\tilde{a} \, a_0}{e^2} \, ,
\eea
where we set the renormalization parameter ${\zeta}=0$, see eq.~(\ref{zeta}).


\section{Solitons with vanishing scalar field}
\label{section-senza-scalare}

In this section we review a class of solutions with vanishing scalar field
and we discuss their phase diagram.
The equations of motions are 
obtained from eq.~(\ref{eom-generiche-2}), setting $H=0$
\beq
 \frac{d}{dr }\le \frac{q}{r} a' \ri = 0 \, ,
 \qquad  
\frac{h'}{h} = 8 \pi G \,   \frac{ {a'}^2}{e^2  \, r}  \, ,
 \qquad
q' = 2 r \, h  \, .
\label{sistemino}
\eeq
Introducing the  integration constant $\tilde{a}$, we find
\beq
 \frac{a'}{r} =   \frac{\tilde{a} }{g \,  (1+r^2)} \, .
\label{def-atilde-da-integration-constant}
\eeq
From eq.~(\ref{def-atilde-da-integration-constant}) we can check that,
if we set $\tilde{a}=0$, then the only solution for $a$
is a constant, $a(r)=a_0$. In this case, there are 
two solutions 
 to the system eq.~(\ref{sistemino}):
\begin{itemize}
\item Global AdS with $a_0=0$, which is required
 in order for the gauge field to be smooth at $r=0$.
\item The BTZ black hole  \cite{Banados:1992wn,Banados:1992gq} 
\beq
h=1 \, , \qquad q= r^2 -\mu \, , 
\label{BTZ-solution}
\eeq
where $\mu=r_h^2$ is related to the horizon radius $r_h$. 
In this case  an arbitrary real value of $a_0$
is allowed, because the singularity in the gauge field is hidden inside the horizon.
\end{itemize}

The matter energy density in eq.~(\ref{densita-energia-generica}) is proportional to the 
square of the $U(1)$ field strength 
 \beq
 \rho=\frac{1}{4} F_{\mu \nu} F^{\mu \nu} 
= \frac{1}{h g} \,  \frac{ \tilde{a}^2}{2  e^2}  \,  \frac{1}{ \,  (1+r^2)}  \, .
\label{rho-u1}
\eeq
For $\tilde{a} \neq 0$, the quantity $\rho$ diverges in correspondence of a horizon,
for which $g(r_h)=0$. This argument shows that our field ansatz with $H=0$ can not
describe any black hole solution with non-zero magnetic field which is smooth outside the horizon.
In the next section we will discuss a class of smooth soliton solution.

\subsection{AdS-Melvin solution}

There is  a solution to the system in eq.~(\ref{sistemino}),
which is the AdS$_3$ analog of the 
 Melvin flux tube in flat space \cite{Melvin:1963qx}.
This background was previously studied in
 \cite{Clement:1993kc,Hirschmann:1995he,Cataldo:1996ue,Dias:2002ps}.
In higher dimensional AdS spacetime, similar solutions were studied in  
\cite{Astorino:2012zm,Lim:2018vbq,Kastor:2020wsm,Anabalon:2021tua,Anabalon:2022aig}.

Following  \cite{Cataldo:1996ue}, we can find a solution to eqs.~(\ref{sistemino})
by performing a double Wick rotation on the charged BTZ solution.
We start from the charged BTZ
black hole, which is described by the following metric
 \cite{Banados:1992wn,Martinez:1999qi}
\bea
&& ds^2 = \frac{d \tilde{r}^2}{f(\tilde{r})} -f(\tilde{r}) d \tilde{t}^2 +\tilde{r}^2 d \tilde{\varphi}^2 \, ,
\qquad
f(\tilde{r})=\tilde{r}^2-r_h^2- 4 \pi G \, \tilde{Q}^2 \log \frac{\tilde{r}^2}{r_h^2} \, ,
\nl
&&A_\mu \, dx^\mu = \tilde{Q} \, \log \le \frac{\tilde{r}}{r_h} \ri \, d\tilde{t} \, ,
\eea
where $\tilde{Q}$ is the electric charge.
The temperature is
\beq
T=\frac{f'(r_h)}{4 \pi} 
=\frac{r_h}{2 \pi } \le 1-\frac{4 \pi G \tilde{Q}^2}{r_h^2} \ri \, .
\label{TTempe}
\eeq
In order for the black hole to have positive temperature,
we need  $r_h^2 \geq  4 \pi G \tilde{Q}^2$.
In this case $r_h$ coincides with the external horizon.
The black hole is extremal for $r_h^2 =  4 \pi G \tilde{Q}^2$.

\begin{figure}[H]
\begin{center}
\includegraphics[scale=0.45]{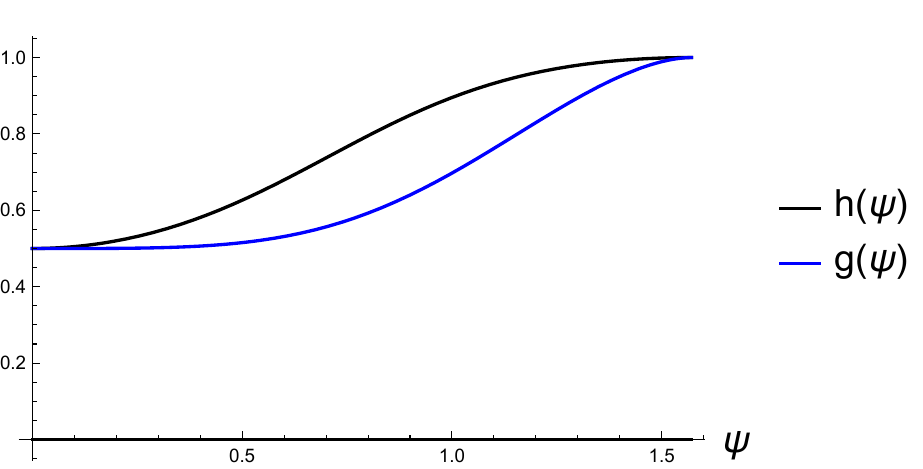} \qquad
\includegraphics[scale=0.45]{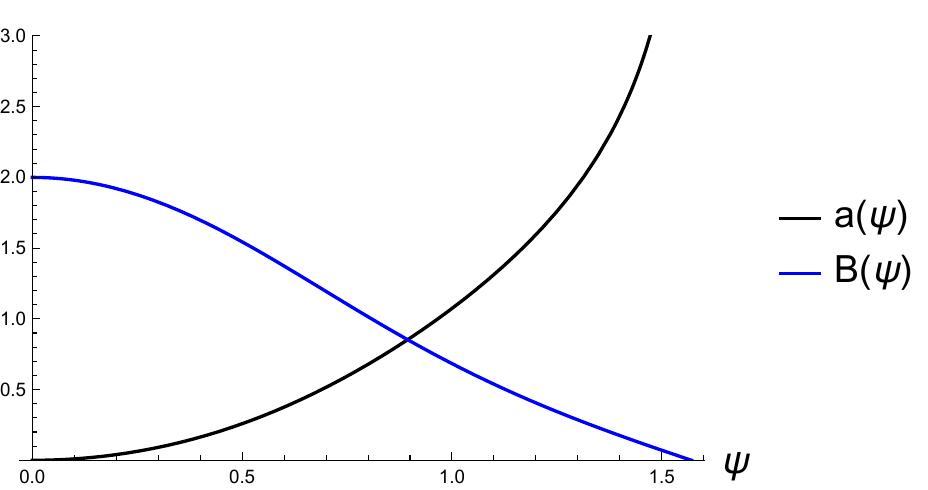}
\caption{  \small The profile functions in eq.~(\ref{soluzione-lambert-W-0}) 
as a function of $\psi$, see eq.~(\ref{coordinata-psi}). 
Here we set $b=1$ and $\frac{e^2}{16 \pi G}=1$.
We plot also the magnetic field $B$.}
\label{nohairlogpsi}
\end{center}
\end{figure}

We perform the double analytic continuation
\beq 
\tilde{t} \to i \frac{\varphi}{2 \pi T} \, , \qquad \tilde{\varphi} \to i \hat{t} \, ,
\eeq
where $T$ is the temperature. The normalization
factor for the analytic continuation of $\tilde{t}$ is needed because
the Euclidean time $\tilde{t}_E$ has periodicity $\frac{1}{T}$ and not $2 \pi$ as  for $\varphi$, i.e.
$T \, d\tilde{t}_E=\frac{1}{2\pi} d \varphi $.
In order to get a real analytically continued solution, we should choose 
 $\tilde{Q}$  imaginary. It is convenient to set $Q=-i \tilde{Q}$.
We finally find
\bea
&& ds^2 =\frac{d \tilde{r}^2}{f(\tilde{r})} -\tilde{r}^2 d \hat{t}^{\,2}  + \frac{f(\tilde{r})}{4 \pi^2 T^2}  \,d\varphi ^2  \, ,
\qquad
f(\tilde{r})=\tilde{r}^2-r_h^2 + 4 \pi G \, Q^2 \log \frac{\tilde{r}^2}{r_h^2} \, ,
\nl
&& A_\mu \, dx^\mu = \frac{ Q}{2 \pi T}  \, \log \le \frac{\tilde{r}}{r_h} \ri \, d \varphi \, .
\label{doppia-continuazione-analitica}
\eea
We will check that the metric of the
 analytically continued solution depends just on the combination
\beq
b=\frac{4 \pi G \, Q^2}{r_h^2} \, .
\eeq
We can express eq.~(\ref{TTempe}) as follows
\beq
\frac{2 \pi \, T}{r_h} =1+b \, . 
\eeq
The relation between the coordinate $r$
in the ansatz eq.~(\ref{metric_generica}) and $ \tilde{r}$ is
given by the size of the transverse $S^1$
\beq
\frac{f(\tilde{r})}{4 \pi^2 T^2} =r^2 \, .
\eeq
We can express $\tilde{r}^2$
in terms of $r^2$ as follows
\beq
\label{soluzione-lambert-W}
\frac{\tilde{r}^2}{r_h^2}=
b \, W_{0} \le \frac{1}{b} 
 \exp \le \frac{ \le 1+ b  \ri^2  r^2+1}{b} \ri \ri
\eeq
where $W_{0}(z)$ denotes the first real branch of the Lambert $W$ function. 
Note  that $r=0$ corresponds to $\tilde{r}=r_h$ and 
that $A_\varphi $ vanishes for this value of $r$.
At large $x$, we can approximate $W_0(x) \approx \log x$.
As a consequence, for large $r$ we have
\beq
\frac{\tilde{r}^2}{r_h^2} \approx r^2 (1+b)^2  = r^2 \, \frac{4 \pi^2 T^2}{r_h^2} .
\eeq
Let us rewrite the metric in eq.~(\ref{doppia-continuazione-analitica})
using the radial coordinate $r$ instead than $\tilde{r}$
\beq
ds^2  =-\tilde{r}^2 d \hat{t}^{\,2}+\frac{16 \pi^2 T^2}{f'(\tilde{r})^2  }  \, dr^2  + r^2 d\varphi ^2 \, .
\eeq
We can use the large $r$
leading behavior $r^2 \approx  \frac{\tilde{r}^2}{4 \pi^2 T^2}$
to fix the normalization of the coordinate $t$ in terms of $\hat{t}$ 
\beq
ds^2  =-\frac{\tilde{r}^2}{4 \pi^2 T^2}  d t^2+\frac{16 \pi^2 T^2}{f'(\tilde{r})^2  }  \, dr^2  + r^2 d\varphi ^2 \, .
\eeq
Comparing this expression with eq.~(\ref{metric_generica}), we find
the relations
\beq
\frac{q}{h}=  \frac{f'(\tilde{r})^2}{16 \pi^2 T^2} \, ,
\qquad
q h=\frac{\tilde{r}^2  }{4 \pi^2 T^2}  \, ,
\eeq
which give
\beq
h=
  \frac{ \frac{\tilde{r}^2}{r_h^2}}{\frac{\tilde{r}^2}{r_h^2} + b } \, ,
\qquad
q = 
 \frac{\frac{\tilde{r}^2}{r_h^2}+ b }{  \le 1+ b \ri^2 }  \, ,
\label{soluzione-lambert-W-0}
\eeq
where $\frac{\tilde{r}^2}{r_h^2}$ is given by eq.~(\ref{soluzione-lambert-W})
in terms of $r$ and the positive real constant $b$.
From eq.~(\ref{doppia-continuazione-analitica}), we find also
\beq
\frac{a }{e}  =
\frac{ 1}{ \sqrt{16 \pi G}} \,  \frac{\sqrt{b}}{1+b} \, \log \le \frac{\tilde{r}^2}{r_h^2} \ri \, .
\label{soluzione-lambert-W-0-a}
\eeq
The solution in eq.~(\ref{soluzione-lambert-W-0}) does not have a horizon
for any values of $b$, because $q$ is always positive.
It also satisfies the condition in eq.~(\ref{no-cone}),
 which is important to avoid conical singularities.
 See figure~\ref{nohairlogpsi} for a plot.

\subsection{Phase diagram}
\label{section-senza-scalare-phases}

The relation between $b$ and 
the integration constant $\tilde{a}$ in eq.~(\ref{def-atilde-da-integration-constant}) is
\beq
4 \pi G \, \frac{\tilde{a}^2}{e^2} 
=\frac{b}{(1+b)^2} \equiv s(b) \, .
\label{eq-b-qtilda}
\eeq
Note that the function $s(b)$ has as maximum $\frac{1}{4}$ for $b=1$.
Let us denote by
\beq
\tilde{a}_c=\frac{e}{\sqrt{16 \pi G} }
\label{atildecritica}
\eeq
  the value of $\tilde{a}$ which corresponds to $b=1$.
 For $\tilde{a}<\tilde{a}_c$
there are two different solutions of eq.~(\ref{eq-b-qtilda})
for $b$ with a given $\tilde{a}$, while for $\tilde{a}>\tilde{a}_c$  there are no solutions for $b$. 
 As a consequence, we find that $\tilde{a}_c$
  is a maximum value for the external current $\tilde{a}$ for the solution 
  in eq.~(\ref{soluzione-lambert-W-0}).

The solution in eq.~(\ref{soluzione-lambert-W-0})
has the following special limits
\begin{itemize}
\item
The value $b=0$ correspond by definition to the AdS 
soliton \cite{Horowitz:1998ha}, which in three dimensions is the same as global AdS$_3$. 
This can be checked as follows.
For large $x$, we can approximate $W_0(x) \approx \log x$.
In this limit, from eq.~(\ref{soluzione-lambert-W}), we find
\beq
\lim_{b \to 0 } \, \frac{\tilde{r}^2}{r_h^2}
 =r^2+1 \, ,
\eeq
which gives, using eqs.~(\ref{soluzione-lambert-W-0}), $h=g=1$.  
\item
At large $b$  the solution   approaches AdS$_3$ in the Poincar\'e patch.
 In this case,  we have that $h=1$ except 
  a small region nearby origin and
 \beq
g\approx \frac{( r^2+ \frac{1}{b}) b^2 }{  \le 1+ b \ri^2 }  \frac{1}{1+r^2} \approx
\frac{r^2+\frac{1}{b}}{r^2+1} \, ,
 \eeq
 which, for $b \to \infty$, is indeed the AdS Poincar\'e patch. 
\end{itemize}

\begin{figure}[H]
\begin{center}
\includegraphics[scale=0.5]{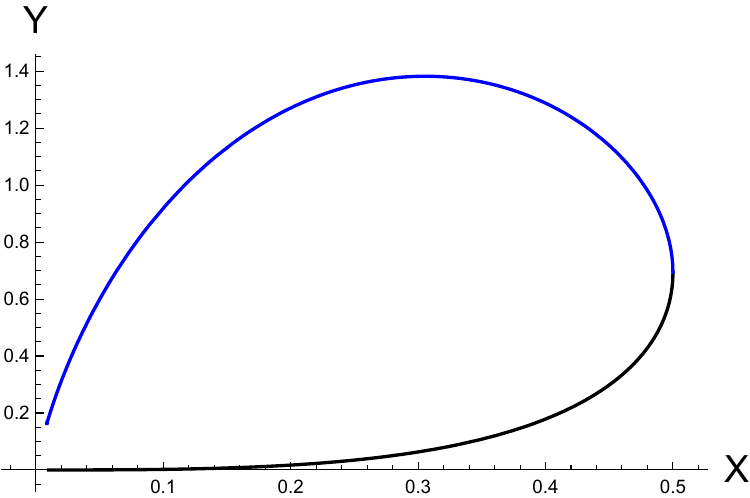} 
\qquad
\includegraphics[scale=0.5]{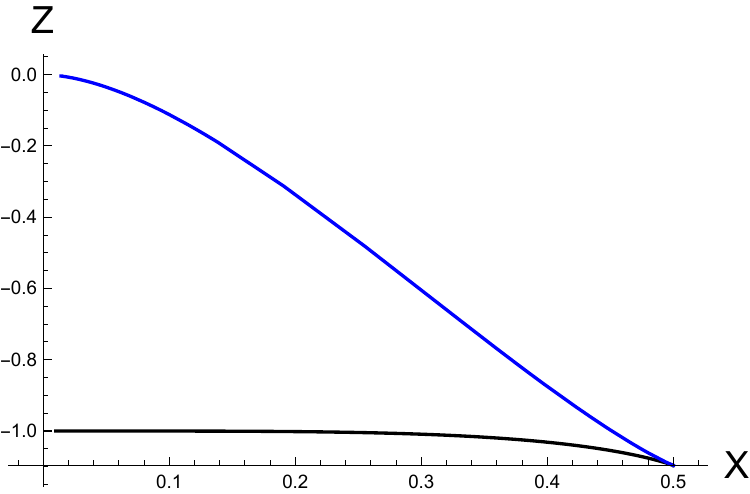} 
\caption{  \small  Left: plot of $Y=\frac{\sqrt{4 \pi G}}{e} \, a_0$ as a function of
$X=\frac{\sqrt{4 \pi G}}{e} \, \tilde{a}=\frac{\sqrt{b}}{1+b}$.
We set for simplicity the scale $r_a=1$.
Right: we plot $Z=16 \pi G \, \mathcal{E}$ as a function of $X$.
The values for solution with $b<1$ and the one with $b>1$ are shown in black and in blue, respectively.}
\label{plotenergysoli}
\end{center}
\end{figure}

  In the left panel of figure~\ref{plotenergysoli} we plot $a_0$
  as a function of the source $\tilde{a}$.
In the right panel of figure~\ref{plotenergysoli} we plot the energy density, 
computed using eq.~(\ref{T-double-trace-massa-minore-3quarti}),
as a function of the source $\tilde{a}$. 
The solution with $b>1$ always has 
higher energy compared to the one with $b<1$
 with the same value of the current.

Global AdS corresponds to the vacuum of a CFT on a circle, with antiperiodic
boundary conditions for the fermions, which give rise to 
 the negative energy due to Casimir effect,
see \cite{Horowitz:1998ha}.
Poincar\'e patch has instead zero energy, because there the dual
CFT has periodic boundary conditions for fermions, which do not break 
supersymmetry. So, we can argue by continuity that the $b<1$
branch corresponds to a phase with antiperiodic boundary conditions for fermions,
while the $b>1$ branch to a phase with periodic ones.
The critical value $\tilde{a}=\tilde{a}_c$ 
is the bifurcation point between the two phases.
 
 There are some similarities with the AdS$_4$ Melvin solution,
  discussed in \cite{Kastor:2020wsm,Anabalon:2021tua}.
   Also in AdS$_4$ there  are two different solutions for the same
 value of the source for the $U(1)$ gauge field
 below the critical value, see \cite{Anabalon:2021tua}. 
 The two branches of the solution, for vanishing $A_\mu$,
  tend to the AdS soliton and to the Poincar\'e patch, respectively.
Note that, in higher dimensions, the source  (with Dirichlet boundary conditions) 
 is the asymptotic value of $ a(r \to \infty)=a_0$,
while in AdS$_3$ we take as source the logarithmic
 falloff of $a$, which is given by $ \tilde{a}$.
In higher dimensions, the Poincar\'e AdS background 
 provides a solution with arbitrary large source $a_0$.
In AdS$_3$  we do not find any solution with 
source $\tilde{a}>\tilde{a}_c$, see eq.~(\ref{atildecritica}).
 


\section{Black holes}
\label{sect-BH-vortex-hair}

In the previous section we have shown that the model 
in eq.~(\ref{modello}) does not admit any
 black hole solutions with a non-zero magnetic field
and vanishing scalar hair.  In this section we will study 
black hole solutions with a non-zero scalar hair.
We find that the Little-Parks periodicity is realized 
in the boundary dual of these black hole solutions.

The black hole horizon is where  the $g(r_h)=0$. 
The location of the horizon $r_h$ is a free parameter of the solution.    
The black hole Hawking temperature $T$ is
\beq
T=\frac{(1+r_h^2) \, g'(r_h)}{4 \pi} \, ,
\label{Temperature}
\eeq
where $\psi_h$ denotes the horizon
in the $\psi$ coordinates in eq.~(\ref{coordinata-psi}).
Assuming that the temperature is not vanishing,
from  eqs.~(\ref{eom-generiche-2}) we find that
we should impose the following conditions at the horizon
in order to avoid singularities
\bea
 \label{H-senza-singo}
&& H'(r_h) = H(r_h) \, \frac{h(r_h)}{g'(r_h)} \, \le
 \frac{(n-a(r_h) )^2}{r_h^2 (1+r_h^2)}
+  \frac{m^2}{1+r_h^2}
\ri \, ,
\nl
&&  a'(r_h)  = 2 e^2 \frac{h(r_h )}{ (1+r_h^2)\, g'(r_h) } \,H^2(r_h)  \, (a(r_h)-n)\, .
\eea
The zero temperature limit needs a more careful analysis, see section~\ref{extremal-limit}.

In condensed matter physics, the Little-Parks effect \cite{Little-Parks}
consists in a periodic behavior of the
 current flowing in a superconducting ring  as a function 
of the magnetic flux which is applied
 in the perpendicular direction. 
 Let us denote by $\varphi$ the compact direction of the cylinder.
 In the Ginzburg-Landau theory, let us denote by
$a_\mu$   the gauge field, by   $D_\mu=\p_\mu-i a_\mu$ the 
covariant derivative, by $v$ the vacuum expectation value
of the order parameter $\phi=v \, e^{i n \varphi}$. 
The superconducting current is proportional to
 \[
 {\rm Im} (\phi^* D_\varphi \phi)= (a_\varphi-n) v^2 \, .
\]
The Little-Parks effect is realized by the invariance
$n \to n+p $ and $a_\varphi \to a_\varphi + p$ with integer $p$.
The same invariance holds in the black hole phase,
 as studied in  \cite{Montull:2011im,Montull:2012fy} for
 the higher dimensional AdS spacetime.
Both the equations of motion  eq.~(\ref{eom-generiche-2})  
and the horizon boundary conditions eq.~(\ref{H-senza-singo}) 
 are indeed invariant under the shift 
\beq
n \to n + p \, , \qquad
a(r)  \to  a(r) + p \, ,
\label{LP-periodicity}
\eeq
where $p$ is an arbitrary integer.

\subsection{Solutions with vanishing external current}

 If we set  $a=n$ as a boundary condition at the horizon,
 from the equations of motion eqs.~(\ref{eom-generiche-2})
 we find that $a$ remains constant.
In appendix~\ref{appe-BH-zero-current} we prove that,
for vanishing external current $\tilde{a}$, 
the profile function $a$ is identically equal to $n$
for the black hole solution. This means that the magnetic flux 
in eq.~(\ref{formula-flusso}) is quantized for the black hole
solution with $\tilde{a}=0$.
This is true even if there is no magnetic field in the solution,
 because the flux has fallen into the singularity, leaving
 a quantized Wilson line outside the horizon.

Specializing eqs.~(\ref{eom-generiche-2}) with $a=n$, we find
\beq
 \frac{d}{dr} \le q \, r  H' \ri = r \, h \,  H   \, m^2   \, , 
\qquad  
 \frac{h'}{h} = 16 \pi G \,   r \, {H'}^2   \, , 
\qquad
 q'= h \, r \, \le 2  - m^2 \, {16 \pi G}  \, H^2 \ri  \, . 
\label{sistema-BH}
\eeq
Note that the system in eq.~(\ref{sistema-BH}),  
 is  independent from the coupling $e$ and from the winding number $n$.

A special class of solutions of eq.~(\ref{sistema-BH}) with vanishing scalar hair is provided by the 
BTZ black hole, 
see eq.~(\ref{BTZ-solution}). 
The mass, the temperature and the entropy  are
\beq
M=2 \pi \, \mathcal{E}=\frac{\mu}{8 G } \, , \qquad T=\frac{\sqrt{\mu}}{2 \pi   } \, , 
 \qquad S=\frac{\pi \sqrt{\mu}}{2 G} \, ,
\label{BTZ-stuff}
\eeq
where $\mu=r_h^2$. The free energy is
\beq
F=M- T S
=-\frac{\pi^2}{2} \frac{1}{G} T^2 \, .
\eeq
There is  a Hawking-Page transition between
global AdS and the  BTZ black hole  \cite{Eune:2013qs} at
 the critical temperature
 \beq
 T_0=\frac{1}{2 \pi } \, .
 \label{HP-temperature}
 \eeq
 
\begin{figure}[H]
\begin{center}
\includegraphics[scale=0.5]{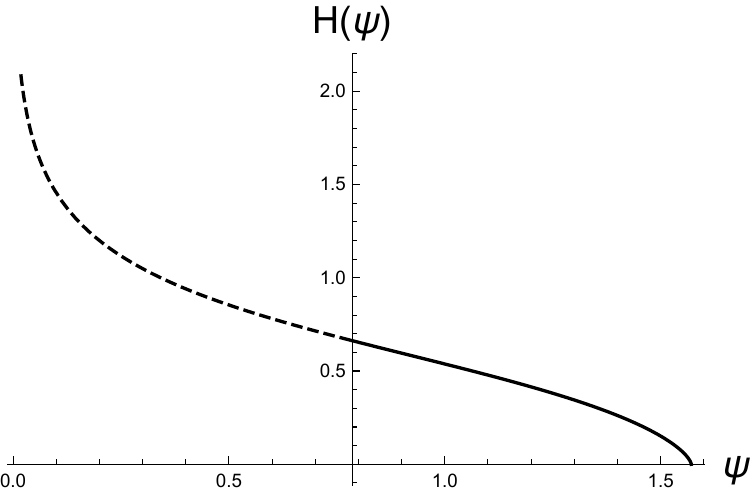} 
\caption{  \small Profile function $H(\psi)$ in eq.~(\ref{soluzione-esatta-BTZ}) for $m^2=-0.9$ and $r_h=1$.
The solid and the dashed parts of the curve correspond 
to the part outside and inside the horizon, respectively. }
\label{HBTZ}
\end{center}
\end{figure}

The BTZ background provides a good approximation  
 nearby the critical temperature for the scalar condensation,
 because $H$ is small in that regime.
Considering a small perturbation of the scalar field $H$ in
 eqs.~(\ref{sistema-BH})  and  neglecting the gravitational backreaction, we find
a linear equation for $H(r)$
\beq
H'' = - \frac{3 r^2-\mu}{r (r^2-\mu) } H' 
+   \frac{  m^2   }{  (r^2-\mu)}
   H \, , 
 \label{Acca-BTZ}
\eeq
which, combined with 
the boundary conditions  in eq.~(\ref{H-senza-singo})
admits the exact solution 
\bea
&&H = \frac{ \Gamma (\sqrt{m^2+1} ) \left(\frac{r^2}{\mu }\right)^{-\frac{\Delta_-}{2}  } 
\, _2F_1\left(\frac{\Delta_-}{2}  , \frac{\Delta_-}{2}  ; \Delta_- ;\frac{\mu }{r^2}\right)}
{\Gamma \left( \frac{\Delta_+}{2} \right)^2}
\nl
&& \quad 
+ \frac{ \Gamma (-\sqrt{m^2+1}  ) \left(\frac{r^2}{\mu }\right)^{- \frac{\Delta_+}{2} } 
\, _2F_1\left( \frac{\Delta_+}{2}   , \frac{\Delta_+}{2} ; \Delta_+ ;\frac{\mu }{r^2}\right)}
{\Gamma \left( \frac{\Delta_-}{2} \right)^2 } \, .
\label{soluzione-esatta-BTZ}
\eea
We show a plot of eq.~(\ref{soluzione-esatta-BTZ})  in figure~\ref{HBTZ}.
The solution  can be extrapolated also 
inside the horizon. Nearby $r=0$, we have a logarithmic singularity in the 
behaviour of the scalar field in eq.~(\ref{soluzione-esatta-BTZ}) 
\beq
H \approx - \frac{2}{\mu^{\frac{\Delta_-}{2}}} \,
 \frac{\Gamma ( \frac{\Delta_+}{2})}{\Gamma (\sqrt{1+m^2}) \,  \Gamma \left( \frac{\Delta_-}{2} \right)}
 \, \log \, r \, .
\eeq
 There is a similar singularity in the black hole solution studied in \cite{Cadoni:2009cq}.

At large $r$, from the exact solution we find  the expansion
\beq
H \approx
 \frac{\Gamma \left(\sqrt{m^2+1}\right)}{\Gamma \left( \frac{\Delta_+}{2} \right)^2}
    \left(\frac{\sqrt{\mu}}{r}\right)^{\Delta_-}
+\frac{\Gamma \left(-\sqrt{m^2+1}\right)}
{\Gamma    \left( \frac{\Delta_-}{2} \right)^2}
 \left(\frac{\sqrt{\mu}}{r} \right)^{\Delta_+} \, ,
\eeq
which corresponds to the double trace coupling
\beq
\kappa
= \frac{   \Gamma \left(-\sqrt{m^2+1}\right) 
\Gamma  \left( \frac{\Delta_+}{2} \right)^2}
   {\Gamma \left(\sqrt{m^2+1}\right) \Gamma \left( \frac{\Delta_-}{2} \right)^2}  \, \le \frac{T}{T_0} \ri ^{2 \sqrt{m^2+1}} \, ,
   \label{kappa-temperatura}
\eeq
where we used $\mu=4 \pi^2 T^2$. The behavior in eq.~(\ref{kappa-temperatura}) follows from dimensional analysis,
using the dimension of the coupling $\kappa$ in the CFT,
see eq.~(\ref{dimensione-kappa}).

In correspondence of the temperature $T_0$ of the Hawking-Page transition,
the value of $\kappa$ in eq.~(\ref{kappa-temperatura})
 is the same as the critical value $\kappa_{c1}$ for $n=0$,
 as we will show in the next section, see $\kappa_v$ in eq.~(\ref{kappa-zero-vuoto}).
 We can interpret eq.~(\ref{kappa-temperatura}) as the critical coupling 
 $\kappa$ for condensation, as a function of temperature.

If we want to include the gravitational backreaction of the scalar $H$
on the metric, we have to solve the system eq.~(\ref{sistema-BH}) numerically,
see appendix~\ref{appe-psi} for details. 
In figure~\ref{esempio-BH-back-vortex} we show an example of solution
with $\kappa=-1$ and $r_h=1$.
Nearby $r \to 0$, the behavior of the solution is
\beq
H \approx - C_H \, \log r \, , \qquad
h \approx h_0 \, r^{16 \pi G \, C_H^2 } \, , \qquad
g \approx g_0 (1-r^2) \, ,
\eeq
for some opportune constants $C_H$, $h_0$ and $g_0$.

  \begin{figure}[H]
\begin{center}
\includegraphics[scale=0.5]{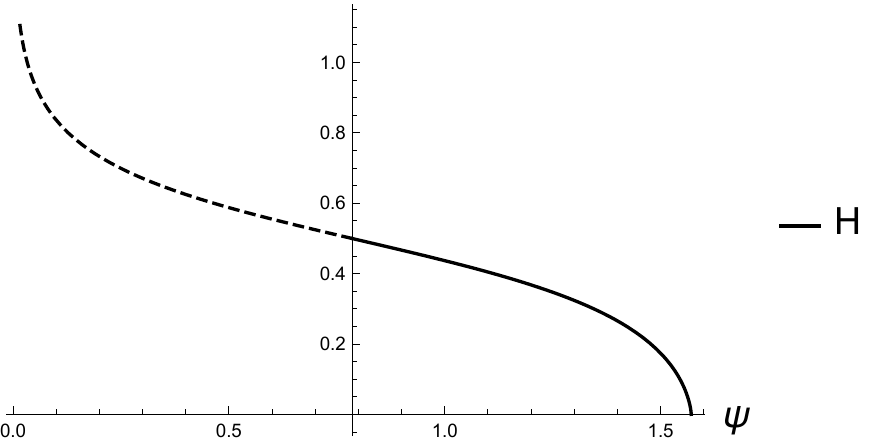} 
\qquad
\includegraphics[scale=0.5]{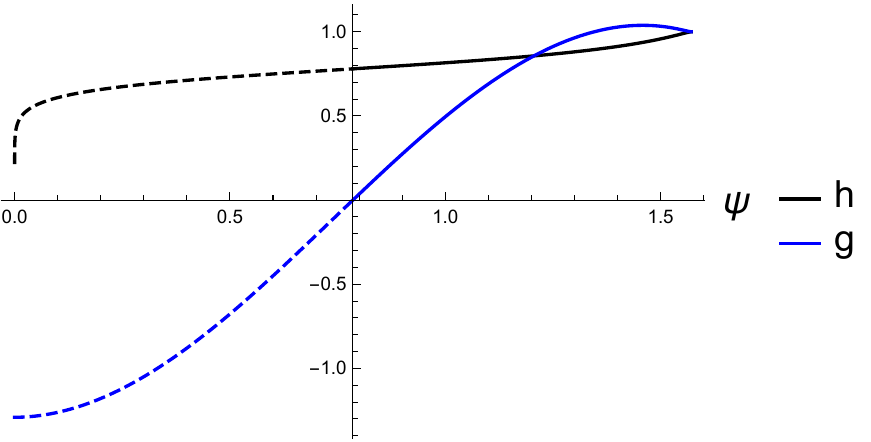} 
\caption{  \small  Example of black hole solution with $\tilde{a}=0$. Here we set $r_h=1$,
 $m^2=-0.9$,  $G=0.1$, $\kappa=-1$.
 The solid and the dashed part of the curve corresponds 
to the part outside and inside the horizon, respectively.
The profile function $a$ is identically equal to the winding number $n$
of the scalar field. The magnetic and electric fields vanish.}
\label{esempio-BH-back-vortex}
\end{center}
\end{figure}

\begin{figure}[H]
\begin{center}
\includegraphics[scale=0.7]{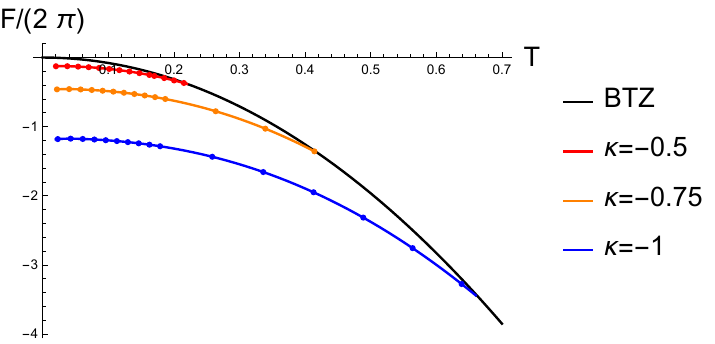} 
\caption{  \small Free energy $\frac{F}{2\pi}$ as a function of temperature $T$
for the hairy black hole solutions with $\tilde{a}=0$, for a few values of $\kappa$.
Here we set  $m^2=-0.9$ and $G=0.1$.
}
\label{FreeEnergyBH} 
\end{center}
\end{figure}

For fixed temperature, in the macrocanonical ensemble the thermodynamically
favored solution is the one with lower free energy
\beq
F= U- T S = 2 \pi \le \mathcal{E} - \, \frac{T \, r_h}{4 \, G}  \ri \, .
\eeq
In figure~\ref{FreeEnergyBH} we show the results of numerical
calculations for the free energy of the 
black hole solution as a function of the temperature, for fixed values of $\kappa$. 
Note that, at low enough temperature, the hairy black hole
has a lower free energy than the BTZ solution. 
From these plots we can check that, at the critical temperature given by
eq.~(\ref{kappa-temperatura}), the hairy solution and the BTZ black hole
have the same free energy.
Above the critical temperature,  the hairy black hole 
with a given $\kappa$ no longer exists.
 
\subsection{Solutions with non-zero external current}

\begin{figure}[H]
\begin{center}
\includegraphics[scale=0.5]{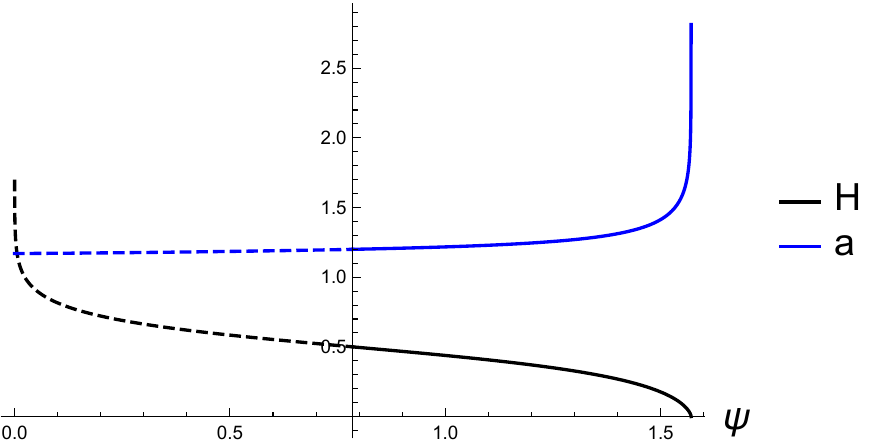} 
\qquad
\includegraphics[scale=0.5]{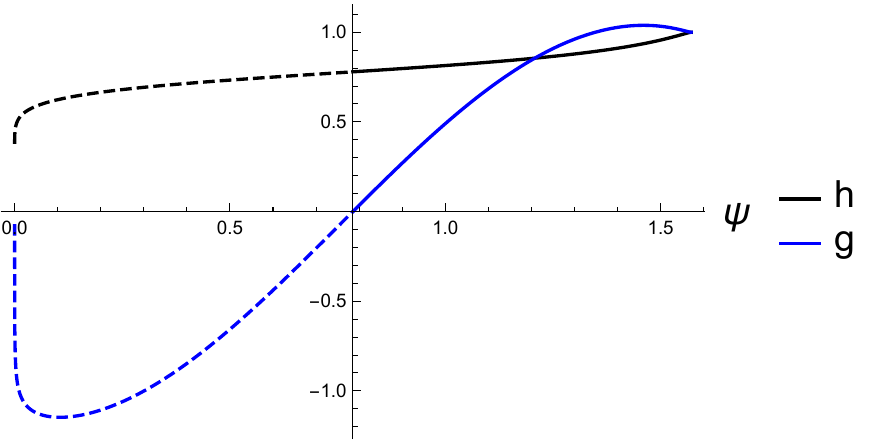} 
\caption{  \small Example of BH solution with non-zero current $\tilde{a}$.}
\label{ConCorrente}
\end{center}
\end{figure}

\begin{figure}[H]
\begin{center}
\includegraphics[scale=0.5]{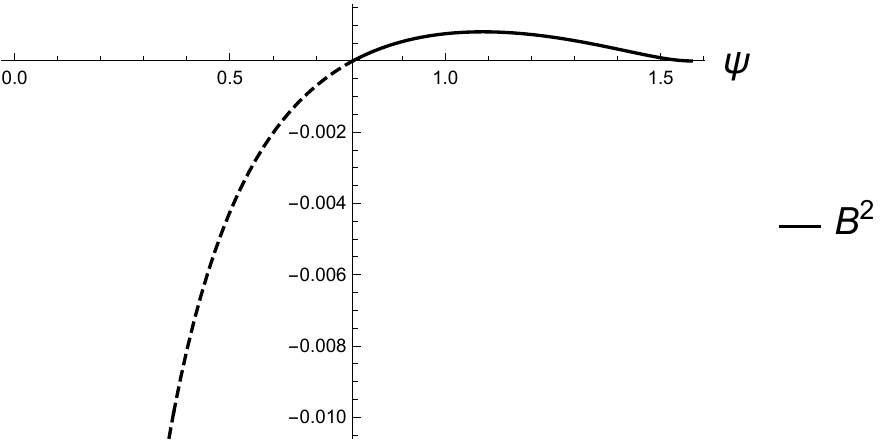} 
\caption{  \small  Plot of $B^2 \propto F_{\mu \nu} F^{\mu \nu}$ 
 for the solution in figure~\ref{ConCorrente}.
Inside the horizon, the magnetic field vanishes and it is replaced by an electric field.
 The solid and the dashed part of the curve corresponds 
to the square of the magnetic field outside and
 to the square of the electric field
 inside the horizon, respectively.
}
\label{ConCorrente2}
\end{center}
\end{figure}

In figure~\ref{ConCorrente} we show an example 
of  numerical solution for the black hole with scalar hair
for non-zero external current $\tilde{a}$. 
In this case, the black solution has a non-zero magnetic field 
outside the horizon.  Note that the magnetic field vanishes inside the horizon.
Inside the horizon, instead there is a non-vanishing electric field
because $r$ becomes a timelike coordinate, see figure~\ref{ConCorrente2}.
In the Ampère law, the term involving the curl of the magnetic field is zero 
inside the horizon and is replaced by the time derivative of the electric field,
 which comes from Maxwell's displacement current term.


\begin{figure}[H]
\begin{center}
\includegraphics[scale=0.55]{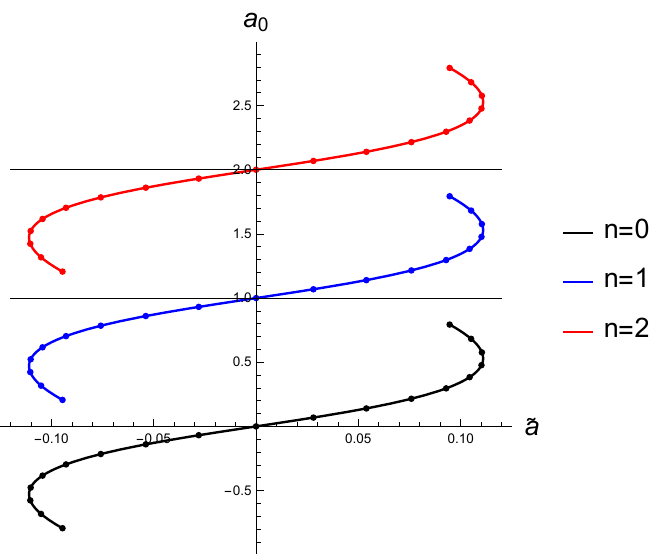} 
\qquad
\includegraphics[scale=0.45]{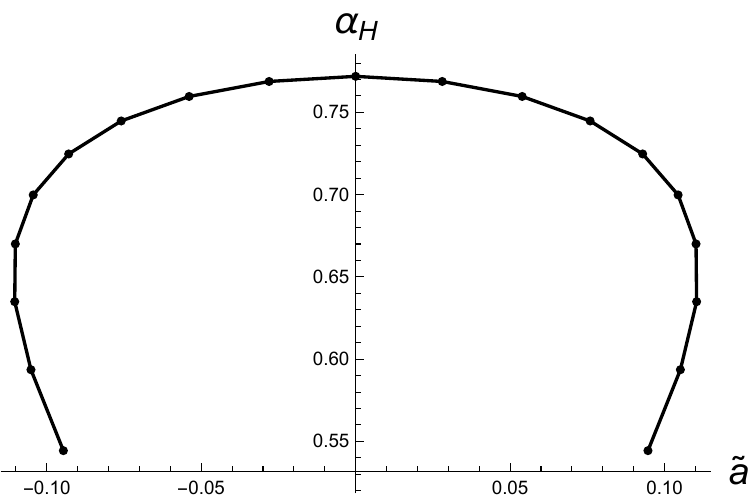} 
\caption{  \small 
On the left panel, we show the expectation value $a_0$ as a function
 of the external current $\tilde{a}$.
For a given winding, we checked that the solution  with lower $|a_0-n|$
has a lower free energy.
On the right panel we plot the condensate $\a_H$ as a function of  $\tilde{a}$.
The presence of a current tends to suppress the VEV of the scalar condensate.
Here we use the numerical values $e=1$, $G=0.1$,   $T=0.2$, $\kappa=-1$.
}
\label{manya}
\end{center}
\end{figure}

In the left panel of figure~\ref{manya} we show $\tilde{a}$ as a function of $a_0$,
extracted from a class of solutions with constant $\kappa$ and $T$.
This illustrates the Little-Parks periodicity in eq.~(\ref{LP-periodicity}).
In the right panel of figure~\ref{manya} we show a plot of the scalar condensate,
which is suppressed by the presence of the external current.
We find that there is a maximum value for $\tilde{a}$, as in the case of vanishing
 scalar field studied in section~\ref{section-senza-scalare}, see eq.~(\ref{atildecritica}).

\subsection{Zero temperature limit}
\label{extremal-limit}

In this section, we will consider the low temperature limit of the
black hole solution with vanishing external current $\tilde{a}=0$.
The zero temperature limit is interesting, because
it is recovered in the transition from the soliton regime,
which we will discuss in section~\ref{section-approaching-black-hole}.
We will find a zero temperature limit which is similar to the one found
for the AdS$_4$ holographic superconductor with $m^2<0$ in  \cite{Horowitz:2009ij}.

Let us first neglect the gravitational backreaction, and solve the equation for $H$
on the zero temperature limit of the BTZ solution,
which is the Poincar\'e patch.
Setting $\mu=0$ in eq.~(\ref{Acca-BTZ}), we find the solutions
\beq
H= A \,  r^{-\Delta_-} + B \, r^{-\Delta_+} \, ,
\eeq
where $A$ and $B$ are integration constants.
For all the values of the integration constants,
the scalar field $H$ diverges at the origin.
We will check that, including the effect of the gravitational
backreaction, the divergence in $H$ becomes milder.

Let us first prove that, in the zero temperature 
limit, the horizon radius (which is defined by the equation
 $g(r_h)=0$) approaches $r_h \to 0$.
From eq.~(\ref{Temperature}), we find that in the zero temperature limit
we have  $g'(r_h)=0$, which is equivalent to $q'(r_h)=0$.
Using   the equations of motion eq.~(\ref{sistema-BH}),
we find that for $r=r_h$ the following quantity should vanish
\beq
h \, r  \le 2  - {16 \pi G}  \, m^2    \, H^2 \ri  =0 \, .
\eeq
Moreover $h(r_h) \neq 0$, otherwise there would be
 a singularity on the horizon.
Because $m^2$ is negative, we have that $r_h$ then should vanish.
In the zero temperature limit, the black hole
approaches a zero entropy field configuration
as in  \cite{Horowitz:2009ij}.

Let us investigate the behavior of the zero temperature limit
of the black hole nearby $r=0$. We will check that $H$ diverges
for $r \to 0$. We can approximate eq.~(\ref{sistema-BH}) as follows
\beq
 \frac{d}{dr} \le q \, r  H' \ri = r \, h \,  H   \, m^2   \, , 
\qquad  
 \frac{h'}{h} = 16 \pi G \,   r \, {H'}^2   \, , 
\qquad
 q'=   - m^2 \, {16 \pi G} \, r \, H^2 \, h \, . 
 \label{sistema-BH-approx}
\eeq
Combining the first and the last equations in the system in eq.~(\ref{sistema-BH-approx}),
we find
\beq
- m^2 \, {16 \pi G}     \, r^2 \, H^2  \, H'  + \frac{q}{h} \,  (H' +  r  H'') = r  \,  H   \, m^2 \, .
\label{approssimazione}
\eeq
Let us assume that in eq.~(\ref{approssimazione})
 the term $ \frac{q}{h}(H' +  r  H'')$ can be neglected (the self-consistency of this assumption can
 be checked at the end of the calculation).
With this approximation, we find
\beq
-  {16 \pi G}     \, r \, H  \, H'   \approx    \, 1 \, ,
\eeq
which gives the approximate solution nearby $r \to 0$
\beq
H \approx \sqrt{\frac{-\log r}{8 \pi G}} \, .
\eeq
From the other equations in eq.~(\ref{sistema-BH-approx}),
 we find
\beq
h \approx \frac{C}{\sqrt{-\log r}} \, ,
\qquad
q\approx  -m^2 \, C \, r^2 \, \sqrt{-\log r } \, ,
\label{singo-metrica}
\eeq
where $C>0$ is an integration constant.
From these expression, we can check that the term $ \frac{q}{h} (H' +  r  H'')  $
can be neglected in eq.~(\ref{approssimazione}).

\begin{figure}[H]
\begin{center}
\includegraphics[scale=0.5]{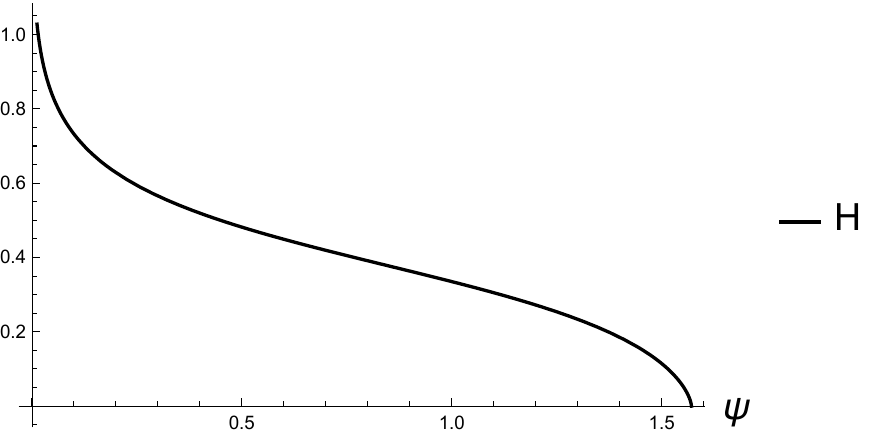} 
\qquad
\includegraphics[scale=0.5]{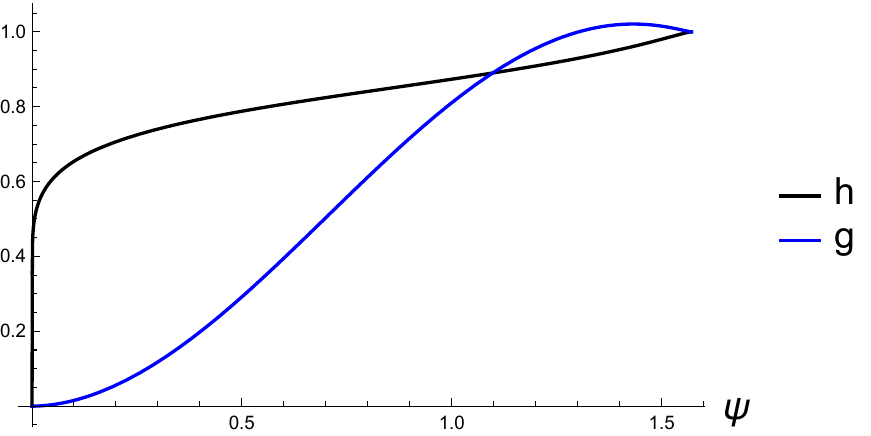} 
\caption{  \small  Zero temperature limit of the black hole solution.
Here we set $m^2=-0.9$,  $G=0.1$, $\kappa=-1$.
 }
\label{esempio-BH-zero-T}
\end{center}
\end{figure}

From eq.~(\ref{singo-metrica}),
 the metric takes the following form nearby $r \to 0 $,
\beq
ds^2 \approx r^2 (-{d \tilde{t}}^2+ d \varphi^2 ) + \frac{dr^2}{ (-m^2) r^2  (- \log r)} \, ,
\eeq
where $\tilde{t}=\sqrt{- m^2} \, C \, t$. 
There is a mild null singularity at $r \to 0$,
which can be detected by the Ricci scalar
\beq
R \approx - 6 m^2 \, \log r \, .
\eeq
We show an example of solution in figure~\ref{esempio-BH-zero-T}.

\section{Solitons with scalar hair}
\label{sec-soliton}

In this section we study soliton solutions with scalar hair
with external current $\tilde{a}=0$.
In section~\ref{exact-sol-soliton} we discuss
an analytical solution  for a probe limit with $G=e=0$,  which
provides a good approximation in the regime of small $\a_H$.
  It turns out that, for a backreacted vacuum solution,
 $\kappa$ is a monotonic function of $\a_H$.
For the vacuum ($n=0$), the exact solution for $G=e=0$  provides
   a good approximation nearby the critical 
$\kappa=\kappa_{c1}$ for the scalar condensation.
 For a vortex with $n \geq 1$,  $\kappa$ instead  is 
 not  a monotonic function of $\a_H$
 and so the solution in  section~\ref{exact-sol-soliton}
does not provide a good approximation at the
  threshold of the scalar condensation. 
  
  For $e,G \to 0$, the equation of motion  for $H$ 
is linear and so $\kappa$ is a constant as a function of $\a_H$. 
 In this limit, the double trace coupling $\kappa$
  depends just on the winding $n$,
thus different windings correspond to different
values of $\kappa$. 
In order to compare the energy of solitons with different 
windings and  with the same
value of the coupling $\kappa$, we need to solve the equations 
including backreaction. In section~\ref{section-numerica-solitone} we 
  discuss some numerical results.
  
 It is interesting that for fixed $G$, $e$ and $n\geq 1$
  there is a 
  second critical value  $\kappa=\kappa_{c2}$ below which 
 no soliton solution can be found.
 For values just above $\kappa_{c2}$, 
 the soliton tends to develop a horizon at $r=r_h$, which is detected
 by the vanishing of $g(r_h)$. 
 We discuss this regime in section~\ref{section-approaching-black-hole}. 
 
 In the limit of large $e$ and fixed $G$, $\kappa$ and $n\geq 1$,
  the magnetic flux is  concentrated in a small region nearby $r=0$.
  We may wonder if  there is a critical value of  $e$ above 
  which the soliton becomes a black hole.
 In appendix~\ref{appe-section-large-e} we present evidences that
this is not the case. Indeed, as  we have seen
 in section~\ref{section-senza-scalare},
no black hole solution can be found with 
vanishing scalar and non-vanishing  magnetic field.


  \subsection{A probe limit in global AdS}
  \label{exact-sol-soliton} 
 
 In this subsection, we discuss a solution of the equations
 of motion  in global AdS,  which is valid
in a probe approximation, for  $G=e=0$,
  in the mass range $-1 \leq m^2 \leq 0$.
 This analytical solution is a useful starting point for the numerical
 analysis carried out in section \ref{section-numerica-solitone}, where we use a relaxation method.
 Moreover, in the case of zero winding $n=0$, this solution 
provides a good approximation at the critical $\kappa=\kappa_{c1}$  for the scalar condensation,
because in this limit the field $H$ is negligibly small and the Maxwell field is zero.

\begin{figure}[H]
\begin{center}
\includegraphics[scale=0.45]{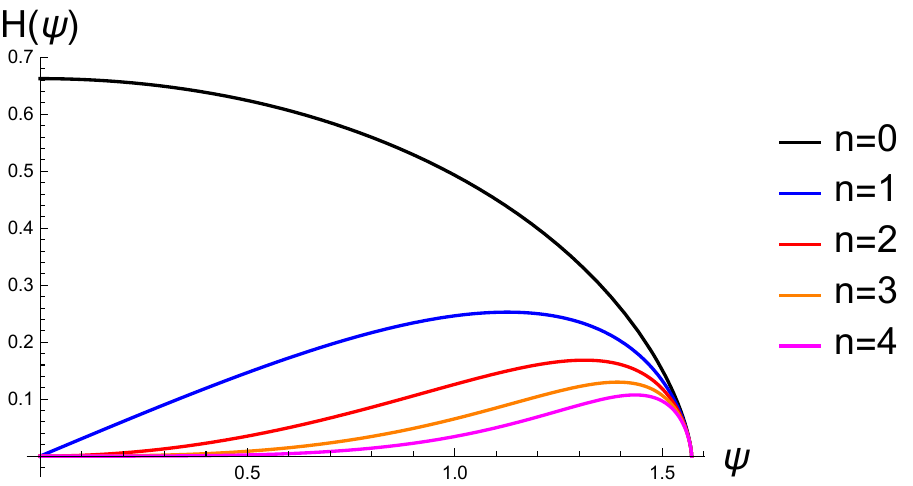}
\qquad 
\includegraphics[scale=0.6]{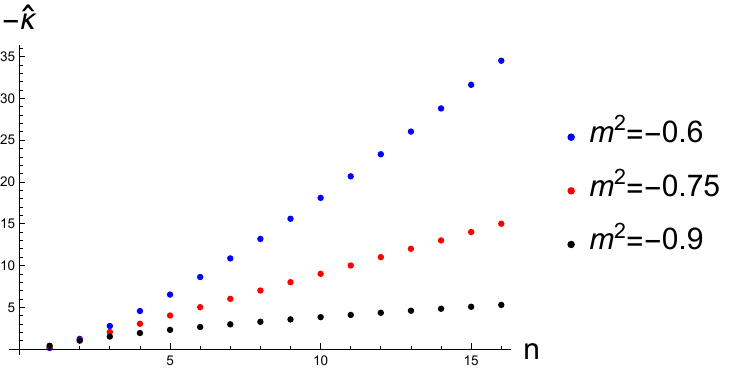} 
\caption{  \small Left panel: solutions $H$ in eq.~(\ref{soluzione-esatta}),
plotted as a function of $\psi=\arctan r$,  for a few values of $n$ and $m^2=-0.9$.
Right panel: Plot of $ -\hat{\kappa}(n)$ for a few values of $m^2$.
}
\label{plotesatte}
\end{center}
\end{figure}

For  $G=e=0$, it is consistent to set $a=0$.
From eq.~(\ref{eom-generiche-2}), we find
 a linear equation for $H(r)$ 
 \beq
H'' = - \frac{1+3 r^2}{r \le 1+ r^2 \ri} H' + \frac{ n^2  +m^2   r^2}{ r^2 \le 1+r^2 \ri} H \, , 
\eeq
which in general has the following exact solution
 \beq
 \label{soluzione-esatta}
H(r)= \frac{ {\Gamma \left(\frac{n+\Delta_+}{2}\right)}^2}{\Gamma\left(\sqrt{1+m^2}\right) \Gamma (n+1)} \,
r^n \, _2F_1\left(\frac{n+\Delta_- }{2},\frac{n+\Delta_+}{2};n+1;-r^2\right) \, ,
 \eeq
normalized with $\a_H=1$.
Plots for different $n$ are shown in the left panel of figure~\ref{plotesatte}.
These solutions correspond to eigenvectors of the laplacian operator
in AdS, see for example \cite{Aharony:1999ti}.

We can express the double trace coupling corresponding to these solutions as
\beq
 \hat{\kappa}(n)=\frac{\b_H}{\a_H}=
\frac{\Gamma\left(-\sqrt{1+m^2}\right)}{\Gamma\left(\sqrt{1+m^2}\right)}
  \left[ \frac{ \Gamma \left(\frac{n+\Delta_+}{2}\right) }{\Gamma \left(\frac{n+\Delta_-}{2}\right)} \right]^2\, .
  \label{kappa-zero}
 \eeq
 Note that the coupling  $ \hat{\kappa}$ is negative,
  as expected in order to achieve spontaneous symmetry breaking.
We can use the approximate formula at large $n$
\beq
 \hat{\kappa} (n) \approx
\frac{\Gamma\left(-\sqrt{1+m^2}\right)}{\Gamma\left(\sqrt{1+m^2}\right)}
\le \frac{n}{2} \ri^{2 \sqrt{1+m^2 }} \, ,
\label{kappa-zero-approx}
 \eeq
which can be derived from the Stirling approximation.
For the special value $m^2 =-\frac{3}{4}$, we have $\hat{ \kappa} (n) \approx -n$.
In the zero mass limit we have $\hat{ \kappa} (n) \propto - n^2$.
In the Breitenlohner-Freedman limit  $m^2 \to -1$, we have  $\hat{ \kappa} (n) \propto - n^0$.
 See the right panel of figure
 \ref{plotesatte} for a plot of $\hat{\kappa}$ as a function of $n$.


\subsection{Numerical solutions }
\label{section-numerica-solitone}

\begin{figure}[H]
\begin{center}
\includegraphics[scale=0.8]{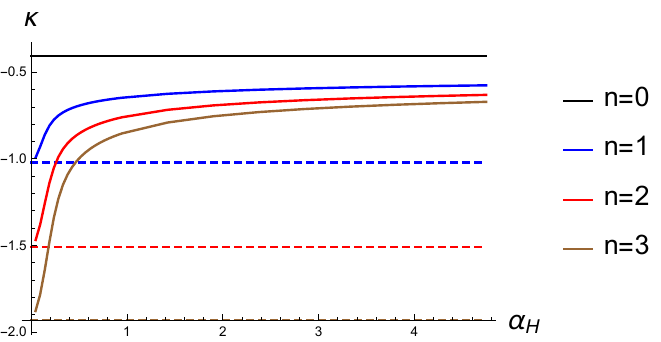}  
\caption{  \small $\kappa$ as a function of $\alpha_H$ 
for the vacuum $n=0$ and for the $n=1,2,3$ vortices. Here we set $m^2=-0.9$ and $e=10$ and 
we neglect the gravitational backreaction, solving the equations of motion on the global AdS metric.
The dotted lines correspond to $e=0$, where $\kappa$  does not depend on $\a$.}
\label{alpha-kappa-noback} 
\end{center}
\end{figure}

In order to study solitonic solutions with non-zero $e$ and $G$,
we have to use numerical methods, see appendix~\ref{appe-psi}.
 If we neglect the gravitational backreaction, the coupling $e$
 by itself does not introduce any non-linearity for 
 the vacuum. In this case $a$ vanishes and  the solution for $H$ is given by
 eq.~(\ref{soluzione-esatta}) with $n=0$ for every value of $e$.
  The multitrace coupling $\kappa$ is independent of $\a_H$
 and is given by $\hat{\kappa}$ in eq.~(\ref{kappa-zero}).
For $n \geq 1$, there is instead a dependence of $\kappa$
on $\alpha_H$, see figure~\ref{alpha-kappa-noback}.
Still, we find that for the vortex configuration $\kappa$ 
is always lower than the vacuum case. Without including gravitational
backreaction,  it is not possible to compare the properties of the 
 vortex solution with a vacuum solution 
with the same choice of double trace coupling $\kappa$.

\begin{figure}[H]
\begin{center}
\includegraphics[scale=0.65]{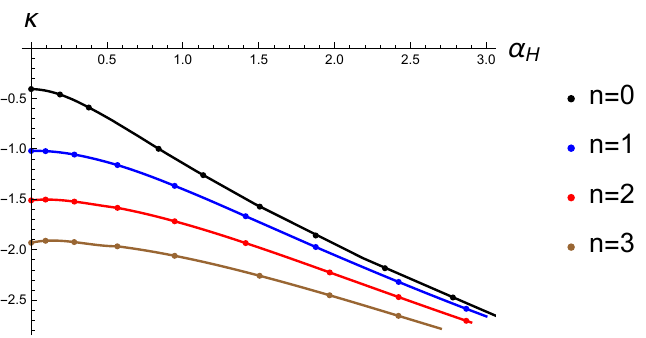} 
\qquad
\includegraphics[scale=0.65]{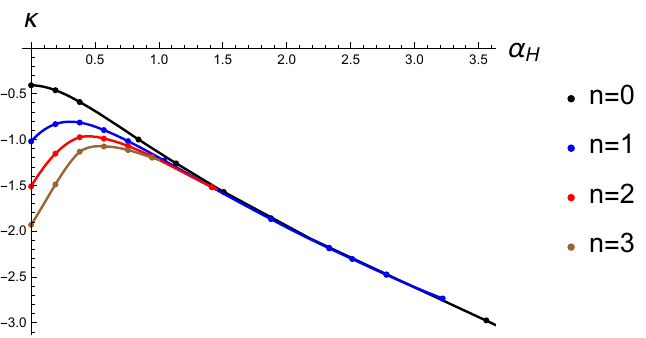} 
\caption{  \small $\kappa$ as a function of $\alpha_H$ 
for the vacuum and $n=1,2,3$ vortices. Here we set $m^2=-0.9$ and $G=0.1$.
In the left panel we set $e=0$, while in the right panel we consider $e=10$.}
\label{alpha-kappa} 
\end{center}
\end{figure}

\begin{figure}[H]
\begin{center}
\includegraphics[scale=0.6]{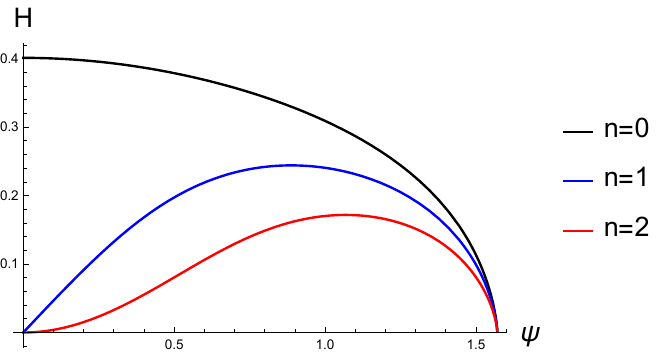} 
\qquad
\includegraphics[scale=0.6]{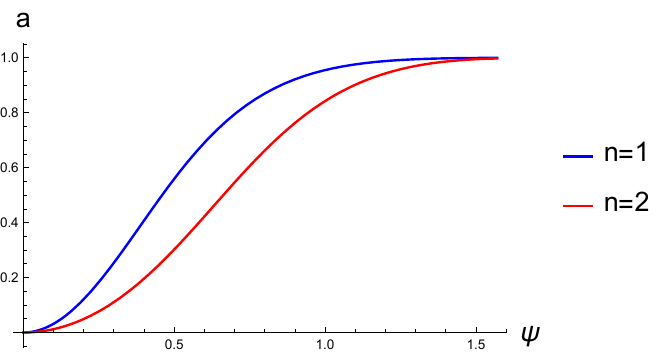} 
\qquad
\includegraphics[scale=0.6]{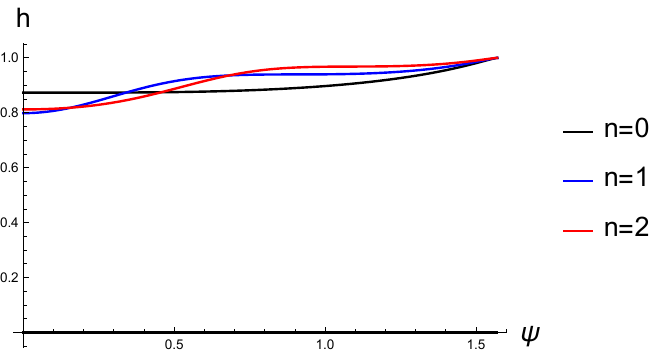} 
\qquad
\includegraphics[scale=0.6]{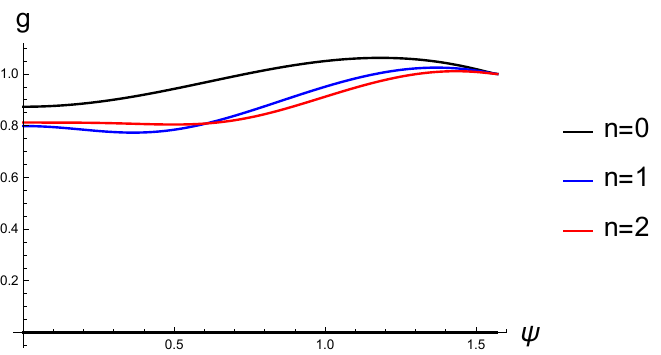} 
\caption{  \small  Soliton solutions  for $n=0,1,2$.
 Here we set $m^2=-0.9$, $e=10$, $G=0.1$ and $\kappa=-1$.
 The three solution have $\a_H=0.840$,  $\a_H=0.738$ and  $\a_H=0.604$, respectively.
We have $\mathcal{E}(n=1)-\mathcal{E}(n=0)=0.301$
 and $\mathcal{E}(n=2)-\mathcal{E}(n=1)=0.307$. 
 }
\label{plot-H-a-hh-g}
\end{center}
\end{figure}

In figure~\ref{alpha-kappa} we show $\kappa$ as a function of $\a_H$
for a class of solitonic vortex solutions with backreaction,
for a few values of the winding number $n$.
Note that $\kappa$, which is always negative,
 has a maximum $\kappa_{c1}$ as a function of $\alpha_H$.
For the vacuum $n=0$
 this maximum corresponds to $\a_H \to 0$ and to 
 \beq
 \kappa_v=\kappa_{c1}(0)= \hat{\kappa}(0)=
 \frac{\Gamma\left(-\sqrt{1+m^2}\right)}{\Gamma\left(\sqrt{1+m^2}\right)}
  \left[ \frac{ \Gamma \left(\frac{\Delta_+}{2}\right) }{\Gamma \left(\frac{\Delta_-}{2}\right)} \right]^2
    \label{kappa-zero-vuoto}
 \eeq
 given in eq.~(\ref{kappa-zero}).
 For $\kappa$ just below $\kappa_{c1}$, we find,
 for $n \geq 1$, two different vortex solutions, with different values of $\a_H$.
 We checked that among these solutions, the one with larger $\a_H$
 has always lower energy, and so it is energetically favored.
 In the following we will always refer to the lower energy solution.
In figure~\ref{plot-H-a-hh-g} we show some examples of solutions.

The magnetic flux $\mathcal{F}$ of the solution 
with vanishing source $\tilde{a}$ is 
\beq
\mathcal{F}=\frac{2 \pi}{e} \,   a(\infty) =\frac{2 \pi}{e} \,  a_0  \, ,
\eeq
 see eq.~(\ref{formula-flusso}).
  While in flat space the flux of a vortex is quantized,
 otherwise its energy is infinite, this is not necessarily the case 
 in the AdS setting studied in this paper, 
 see the discussion nearby eq.~(\ref{stima-energetica}).
 Note that the magnetic flux instead is quantized for the black hole
 solutions with $\tilde{a}=0$, as we proved in appendix~\ref{appe-BH-zero-current}.

In figure~\ref{PlotFlux} we plot the vortex flux as
a function of the scalar expectation value $\a_H$.
We see that $a_0$ tends to approach the winding number $n$ for large 
enough $\a_H$, which correspond to large enough
$|\kappa|$. In this limit, the quantization of the flux holds
in a similar way as in flat space.
Another limit in which the quantization of the flux holds
is the one of large $e$ with fixed $\kappa$.
\begin{figure}[H]
\begin{center}
\includegraphics[scale=0.7]{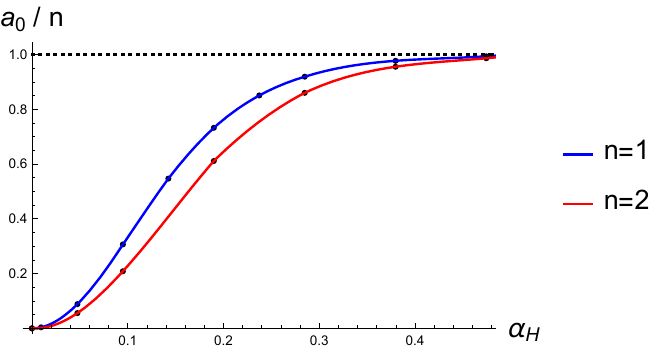}  
\caption{  \small The normalized flux $\frac{a_0}{n}$ as a function of $\a_H$ for the $n=1$ and  $n=2$ vortices.
Here we set $e=10$, $m^2=-0.9$ and  $G=0.1$.
}
\label{PlotFlux} 
\end{center}
\end{figure}

\subsection{Approaching the critical solution}
\label{section-approaching-black-hole}

Let us fix $e$ and $G$ and decrease the negative coupling $\kappa$.
From figure~\ref{alpha-kappa}, we see that  $\a_H$ 
increases.\footnote{Here we restrict, for a given $\kappa$, to the solution
with higher $\a_H$, which turns out to be more stable because 
it always has a lower energy.}
As a consequence, the scalar field  $H$  in the interior of the geometry
also tends to increase. For $n\geq 1$, the magnetic field is localized in a increasingly
narrow region nearby $r=0$.
Intuitively, if some energy is concentrated inside a region
whose size approaches the Schwarzschild radius, 
a black hole horizon will form.
For this reason, we expect that it exists a critical value of the coupling $\kappa$,
which we denote by $\kappa_{c2}$, above 
which no soliton solution with a given winding $n\geq 1$ can exist,
due to black hole horizon formation.
A similar transition between monopole solutions and black holes was  investigated 
in flat space  in \cite{Lue:1999zp,Weinberg:2001gc}.

In order for the critical solution to exist,
we need that, for a certain value of $r=r_0$, the function
$q(r)$ has a positive minimum $q(r_0)$ which tends to zero,
in  such a way that $q'(r)<0$ for $r<r_0$.
The minimum condition $q'(r_0)=0$,
with the horizon condition $q(r_0) = 0$,
tells us that the black hole should have vanishing temperature,
see eq.~(\ref{Temperature}).
In section~\ref{extremal-limit} we have found  that the
zero temperature condition implies
that the radius $r_h$ of the horizon of the black hole vanishes,
and so that the entropy is zero.
The zero temperature limit 
of the black hole solution
corresponds to 
a zero entropy configuration, where the horizon
lies on top of a mild logarithmic curvature singularity
at $r_h=0$.

Let us inspect the equation for $q$ 
in eq.~(\ref{eom-generiche-2})
\beq
q'= h \le 2 r - {16 \pi G} \le \frac{(a-n)^2}{r} + m^2  r \ri  \, H^2 \ri \, . 
\eeq
From this expression, we find $q'(r)>0$ 
 if $m^2$ is negative and $n=0$, because in this case $a=0$.
No critical solution can exist for $n=0$.
As a consequence,  for $n=0$ a soliton solution
 without horizon always exists for every value of $\a_H$.
 This is consistent with the numerical solutions.
Instead, for higher windings $n \geq 1$  we can always find a region,
nearby $r=0$, where
\[
\frac{(a-n)^2}{r^2}    > - m^2 \, ,
\]
in such a way that $q'$ might become negative.
Critical solutions may then exist for $n \geq 1$. 

\begin{figure}[H]
\begin{center}
\includegraphics[scale=0.5]{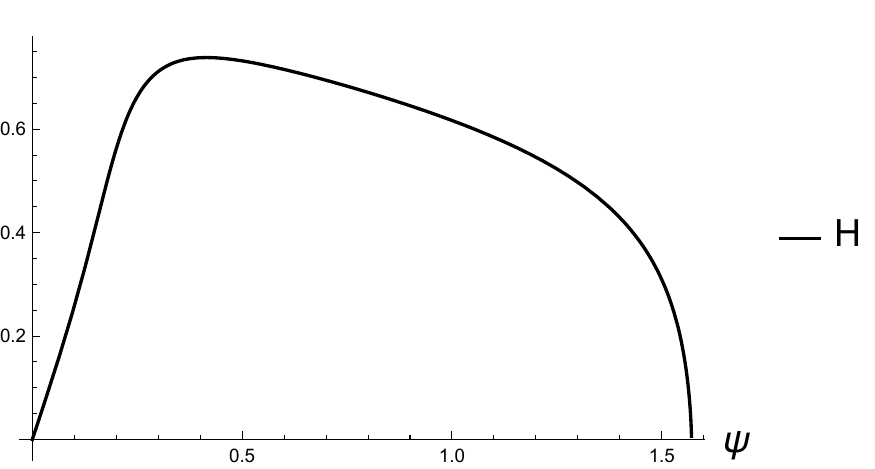} 
\qquad
\includegraphics[scale=0.5]{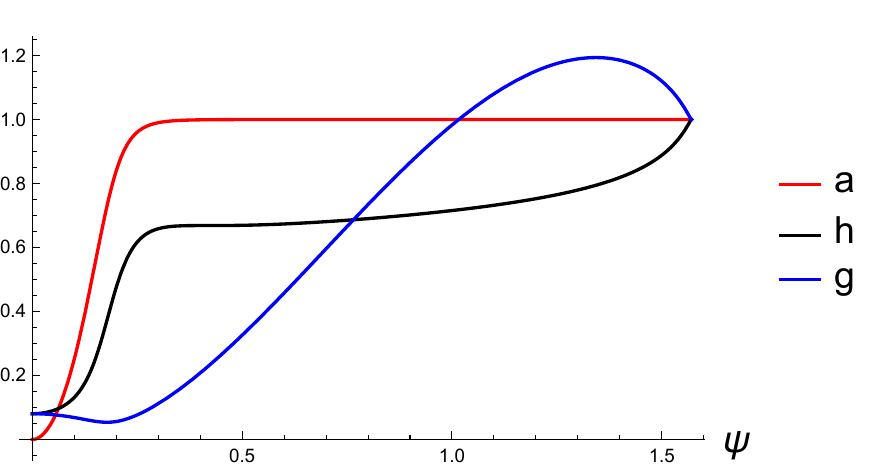} 
\caption{  \small  Example of nearly critical soliton solution with $n=1$.
Here we set $e=10$, $m^2=-0.9$,  $G=0.1$, $\kappa=-2.735$,
$\a_H=3.220$.
 }
\label{near-critical}
\end{center}
\end{figure}

In order to give a criterium for the formation of a horizon,
 let us integrate
 eq.~(\ref{constraint-specializzata})
from $r=0$ to the would be horizon $r_0$.
 Assuming that $q$ has a positive minimum at $r=r_0$, we find 
\beq
 16 \pi G \int_0^{r_0} \rho \, r \, dr \leq r_0^2 +1  
\label{criterio-formazione-orizzonte}   \, .
\eeq
If the integral of $\rho$
inside a ball with radius $r_0$ saturates
the bound (\ref{criterio-formazione-orizzonte}), a horizon forms.
See figure~\ref{near-critical} for an example of nearly critical solution 
for $n=1$.  We checked numerically that 
 eq.~(\ref{criterio-formazione-orizzonte}) tends to be saturated
 for the critical solution.

For fixed $e$ and $G$, a vortex soliton 
with a given $n \geq 1$ exists just
in a finite window of multitrace coupling $\kappa$.
There is an upper bound given by $\kappa_{c1}$, see figure~\ref{alpha-kappa}. 
There is also a lower bound given by $\kappa_{c2}$,
otherwise the soliton tends to develop a horizon at $r_h \to 0$, 
as shown in figure~\ref{near-critical}. For $e=10$ and $G=0.1$, we 
have  $\kappa_{c1} = -0.8$ and $\kappa_{c2} = -2.75$.

In appendix~\ref{appe-section-large-e} we briefly consider
the large $e$ limit with fixed $\kappa$. 
In this case, we do not 
find any critical $e$ for which 
the soliton solution tends to develop a horizon.


\section{Conclusions}
 \label{sec-conclusion}
 
\begin{figure}[H]
	\begin{center}
		\includegraphics[scale=0.75]{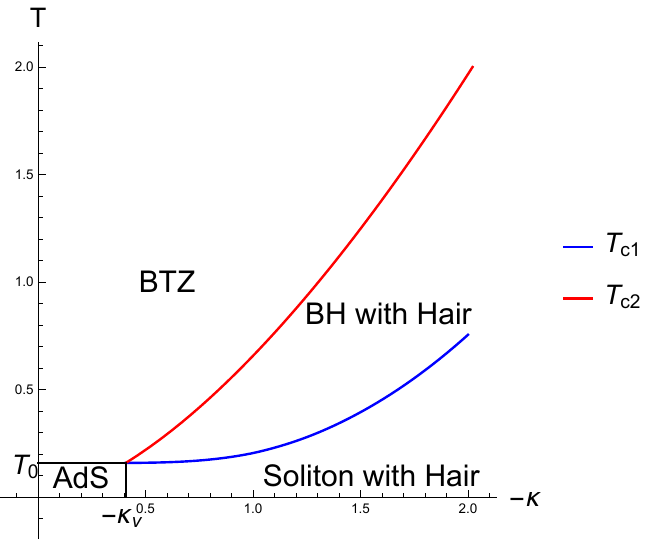} 
		\caption{  \small Phases diagram for $\tilde{a}=0$ in the $(-\kappa,T)$ plane for
			for the vacuum solution $n=0$, for $m^2=-0.9$ and $G=0.1$.
		}
		\label{phases-diagram-0}
	\end{center}
\end{figure}

In this paper, we studied various soliton and black hole
solutions in AdS Einstein gravity with a Maxwell and a charged scalar 
field.  Here we summarize our  results about the phase diagram with  vanishing external current $\tilde{a}$
as a function of the temperature $T$, the double trace coupling $\kappa$,
and the winding number $n$.
At low enough temperature, the soliton solution 
can have lower free energy compared to the black hole.
This gives rise to a first order phase transition,
which is analog to the one studied by
Hawking and Page  \cite{Hawking:1982dh}.
Moreover, the double trace coupling $\kappa$ can trigger
the condensation of the scalar order parameter,
giving rise to a second order phase transition
to a superconducting phase \cite{Faulkner:2010gj}.
As in \cite{Nishioka:2009zj,Horowitz:2010jq}, we can have a total of four possible phases,
corresponding to a smooth soliton and black hole
 with or without the presence of scalar hair.
 For zero winding number $n=0$ the phase diagram 
 is independent of the bulk $U(1)$ gauge coupling $e$,
 because the solutions have vanishing field strength $F_{\mu \nu}$.
 In this case, a soliton   exists for every value
  of $\kappa < \kappa_v$, see eq.~(\ref{kappa-zero-vuoto}).
 We show  the phase diagram
  in figure~\ref{phases-diagram-0}.
     For $\kappa \geq \kappa_v$, there is no scalar condensation
 and there is the Hawing-Page first order phase transition
 at the temperature $T_0=\frac{1}{2 \pi}$.
 For  $\kappa < \kappa_v$, the hairy soliton solution has a lower free 
 energy at low enough temperature. At a  critical temperature $T_{c1}$,
 which depends on the Newton constant $G$ and that
  can be found numerically by comparing the free energy
 of the black hole and of the soliton phase,
 there is a first order transition to the hairy black hole phase.
 From numerical calculations, we find that $T_{c1}$
is an increasing function of $|\kappa|$, see figure~\ref{phases-diagram-0}.
At a higher temperature $T_{c2}$, 
which is independent of $G$ and that 
can be found by solving eq.~(\ref{kappa-temperatura}) to be
\beq
  T_{c2} = T_0
\left[ \frac{\Gamma \left(\sqrt{m^2+1}\right) \Gamma \left( \frac{\Delta_-}{2} \right)^2}
 {   \Gamma \left(-\sqrt{m^2+1}\right) 
\Gamma  \left( \frac{\Delta_+}{2} \right)^2}
  \, \kappa \right]^{\frac{1}{ 2 \sqrt{m^2+1}}} \, ,
  \label{tempe-c2}
\eeq
 there is a second order  phase transition 
 to a non-superconducting phase,
which is described by the BTZ black hole solution. 
The fact that for $\kappa=\kappa_v$ the
critical temperature $T_{c2}$ in eq.~(\ref{tempe-c2}) is equal to $T_0$
provides evidence for the quadruple point in 
the phase diagram in figure~\ref{phases-diagram-0}.  
This is a different feature compared to \cite{Nishioka:2009zj,Horowitz:2010jq},
where the quadruple point can be realized just for 
a fine-tuned choice of the couplings.

\begin{figure}[H]
	\begin{center}
		\includegraphics[scale=0.55]{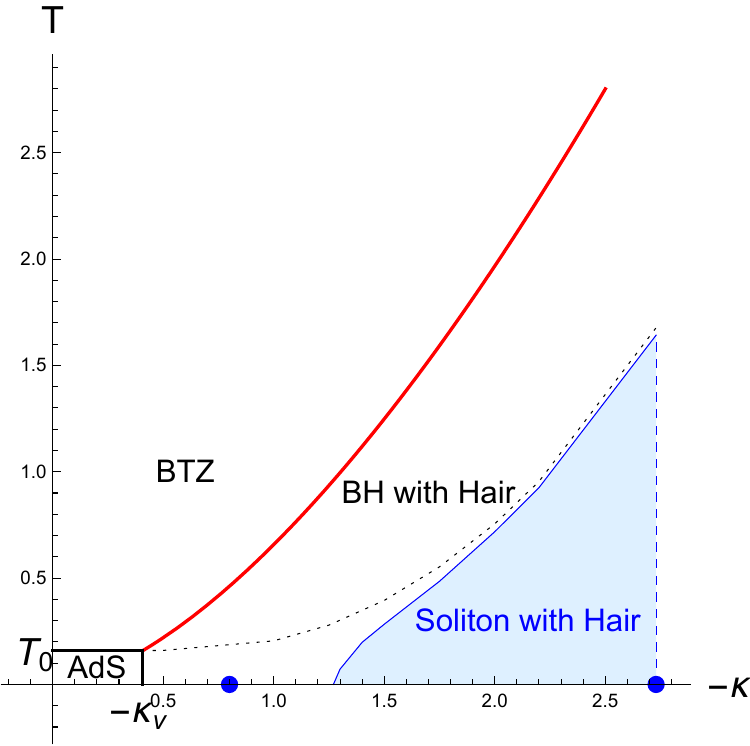}
		\caption{  \small   \small
			Phase diagram for the vortex solution with $n=1$, 
			for $m^2=-0.9$, $G=0.1$, and  $e=10$.
			The two blue points mark the
			values  $\kappa_{c1} $ and $\kappa_{c2}$,
			which correspond to the  maximum and minimum
			value of $\kappa$ for which a solitonic 
			vortex solution exists.
			The solid blue line corresponds to the first order phase transition
			between soliton and black hole,
			due to the change of the minimum of the free energy 
			(for comparison, the same transition
			is shown for $n=0$ in a black dotted line).
			The dashed blue line is the border of the region
			where the soliton with $n=1$ does not exist, because 
			a black hole horizon tends to form.
		}
		\label{phases-diagram-1}
	\end{center}
\end{figure}

For non-zero winding $n$, 
 a soliton solution  exists just in a specific 
 window of the double trace coupling $\kappa_{c2} \leq \kappa \leq \kappa_{c1} $
 which depends both on the couplings $G$, $e$ and on the winding $n$.
For example,  for $n=1$ and for the value of the couplings $e=10$ and $G=0.1$ we have 
 $\kappa_{c1} = -0.8$ and $\kappa_{c2} = -2.75$.
  We show the phase diagram
 in figure~\ref{phases-diagram-1}.
For a value of $\kappa$ below $\kappa_v$
and outside the window $\kappa_{c2} \leq \kappa \leq \kappa_{c1} $,
there are just two phases, described by the hairy black hole
and by the BTZ solution. Instead, in the window
$\kappa_{c2} \leq \kappa \leq \kappa_{c1} $, there is also 
the soliton vortex phase. In the low temperature 
limit, it can happen that the vortex has a lower free energy than the black hole,
see the blue region in figure~\ref{phases-diagram-1}.

In the case of vanishing scalar hair,
we also studied the phase diagram
with non-zero external current $\tilde{a}$,
see section \ref{section-senza-scalare}.
For a given value of $\tilde{a}$ below a critical value  $\tilde{a}_c$ given  in eq. 
(\ref{atildecritica}), we find that there are two solutions.
For vanishing $\tilde{a}$ the two solutions correspond
to global AdS and to the Poincar\'e patch. For the critical value
$\tilde{a}_c$  the two branches of solutions coincide, see figure \ref{plotenergysoli}.
We find that the branch which is continuously connected
to global AdS has always a lower energy
compared to the branch which is continuously connected to the Poincar\'e patch.
An analogous  phase diagram has been investigated in higher 
dimensions in  \cite{Kastor:2020wsm,Anabalon:2021tua}.

In the hairy black hole phase, we have found
by numerical methods that
there is also a maximum value of the current $\tilde{a}$,
see figure \ref{manya}. This is consistent with the presence
of  the Little-Parks periodicity, which is  is realized
by the shift symmetry in eq.~(\ref{LP-periodicity}), as in  \cite{Montull:2011im,Montull:2012fy}. 
In the soliton phase, instead, if we use regular boundary conditions at $r=0$,
 we find no evidence of Little-Parks periodicity. 
Violations of  LP periodicity has been studied in the condensed matter literature, 
for example    in small cylinders, when the size of the system 
is of the same order of magnitude of the coherence length
of the superconductor, see \cite{Vakaryuk,Loder,Wei}.  




\subsection*{Acknowledgments}

The work of R.~A., S.~B., and G.~N. is supported by the INFN special research project
grant ``GAST'' (Gauge and String Theories). 
The work of N.~Z. is supported by MEXT KAKENHI Grant-in-Aid for Transformative Research Areas A “Extreme Universe” No. 21H05184.

 \section*{Appendix}
\addtocontents{toc}{\protect\setcounter{tocdepth}{1}}
\appendix

\section{Numerical methods}
\label{appe-psi}

In order to find numerical solutions to the equations of motion,
it is useful to introduce the compact coordinate $\psi = \arctan r$. 
In this coordinates the metric of global AdS is
\beq
ds^2=\frac{1}{\cos^2 \psi} \le - \, h g \, dt^2 +\frac{h}{g} d \psi ^2 +\sin^2 \psi  \, d \varphi^2 \ri \, .
\eeq
The equations of motion eq.~(\ref{eom-generiche-2}) in the $\psi$ variable read
\beq
\begin{aligned}
E_H &=H'' - \frac{1}{g} \left[ \le (n-a)^2 \csc^2 \psi +m^2  \sec^2 \psi \ri h \, H 
- \le g \, \csc \psi \, \sec \psi + g' \ri H' \right] =0 \, , \\
E_a &=a'' - \frac{1}{g} \left[ 2 e^2  (a-n) \, h \, H^2 \, \sec^2 \psi + \le g \, \csc \psi \, \sec \psi - g' \ri a' \right]=0 \, , \\
E_h &= h' - {8 \pi G} \, h \, \cos \psi \le \cos \psi \cot \psi \, \frac{ {a'}^2}{e^2 } +2 \sin \psi \, {H'}^2 \ri =0 \, , \\
E_g &= g' - \left[ 2 \tan \psi \le h-g \ri - {16 \pi G} \, h \, \tan \psi \le (n-a)^2 \cot^2 \psi +m^2 \ri H^2 \right] = 0\, ,
\end{aligned}
\label{eqs-backreaction-psi}
\eeq
where $'$ denotes the derivative with respect to $\psi$.

In the case of the black hole with scalar hair,
we can find numerical solutions 
by solving directly eqs.~(\ref{eqs-backreaction-psi})
 as a system of ODE, with boundary  conditions
 at the horizon $\psi_h=\arctan(r_h)$.
In terms of the $\psi$ coordinate, the horizon boundary
conditions in eq.~(\ref{H-senza-singo}) are
\bea
 \label{senza-singo-psi}
&& H'(\psi_h) = H(\psi_h) \, \frac{h(\psi_h)}{g'(\psi_h)} \,
\le  \frac{(n-a(\psi_h) )^2}{\sin^2 \psi_h}
 +\frac{m^2}{\cos^2 \psi_h} \ri \, ,
 \nl
&& a'(\psi_h)  =
  2 e^2 \frac{h(\psi_h )}{\cos^2 \psi_h \, g'(\psi_h) } \,H^2(\psi_h)  \, (a(\psi_h)-n)\, .
\eea 

 In the case of the soliton, it is a little subtle to solve the
system of ordinary differential equations
in eq.~(\ref{eom-generiche-2}) numerically. 
The difficulties basically arise because the smooth solutions for $H$ and $a$
are  fine-tuned as functions of boundary conditions.
We solve the equations for $H$, $a$ with fixed $g,h$,
using the relaxation method.
 We add an auxiliary time dependence to the profile functions $H(\psi,t)$
 and $a(\psi,t)$ and we solve for the partial differential equations (PDE) system
 \beq
  \frac{\partial H}{\partial t}= E_H \, ,
  \qquad
 \frac{\partial a}{\partial t}= E_a \, ,
 \label{PDE-ausiliaria}
 \eeq
where $E_H$ and $E_a$ are in eq.~(\ref{eqs-backreaction-psi}).
The vanishing of $E_H$ and $E_a$ is equivalent to the equations of motion
for $H$ and $a$.  We use the exact solution 
with $e=G=0$ in eq.~(\ref{soluzione-esatta}) as an initial condition
at $t=0$ for the PDE system.
For the scalar $H$, we impose the boundary condition with fixed $\a_H$
 at $\psi=\frac{\pi}{2}$,   where $\a_H$ can be computed as follows
 \bea
 \a_H 
 = \lim_{\psi \to \frac{\pi}{2}} \frac{(\tan \psi)^{1-2 \sqrt{1+m^2}}}{2 \sqrt{1+m^2}} \, 
  [ H(\psi) (\tan \psi)^{\Delta_+}]' \cos^2 \psi \, .
 \eea
Then, we can extract $\b_H$ by performing a  fit on the tails of the solution. 
In order to study the vortex solution without gravitational backreaction,
we set $h=g=1$, and we solve the PDE system in eq.~(\ref{PDE-ausiliaria}).

In order to find solutions which include  gravitational backreaction,
we can first solve the ordinary differential equations (ODE)
for the metric functions $h$ and $g$
in eq.~(\ref{eqs-backreaction-psi})
 for fixed $H$ and $a$, given by the solution of
 eqs.~(\ref{PDE-ausiliaria}) on global AdS background.
We can then iterate the procedure several times,
 each time  solving first the PDE for $H$ and $a$,
and then again solving the ODE for $h$ and $g$.
Each time we solve the ODE for $h,g$, we can 
monitor $E_H$ and $E_a$ for convergence. 
We continue until the procedure converges and  $E_H$ and $E_a$ 
are small enough. When this procedure converges, we can find 
backreacted solutions with arbitrary accuracy.

\section{Proof that $a(r)=n$ for a black hole with zero external current }
\label{appe-BH-zero-current}

Here we prove that, for the black hole solution with zero external current $\tilde{a}=0$,
 we can set $a(r)=n$. 
 The proof is analogous to the one of
  the scalar no-hair theorem \cite{Bekenstein:1971hc,Weinberg:2001gc}.
 Let us suppose that it is not the case and then we will find a contradiction.
 Let us start from the second equation in the system in eq.~(\ref{eom-generiche-2}) 
and  multiply both sides by $a-n$ 
\beq
(a-n) \frac{d}{dr} \le  \frac{g \,  (1+r^2)}{r} a'  \ri =2 e^2 \frac{(a-n)^2}{r} \, h\,H^2 \, .
\eeq
We can write the above equality as
\beq 
 \frac{d}{dr}  \left[ (a-n)  \le  \frac{g \,  (1+r^2)}{r} a'  \ri \right]={a'}^2  \le  \frac{g \,  (1+r^2)}{r} \ri 
 +2 e^2 \frac{(a-n)^2}{r} \, h \, H^2 \, .
\eeq
By denoting the horizon radius  $r_h$ as the most external zero of $g(r)$,
we integrate
\beq 
 \int_{r_h}^\infty  \frac{d}{dr}  \left[ (a-n)\le \frac{g \,  (1+r^2)}{r}  a' \ri \right] dr
  = \int_{r_h}^\infty \left[ {a'}^2  \le \frac{g \,  (1+r^2)}{r}  \ri  
 +2 e^2 \frac{(a-n)^2}{r} \, h \, H^2 \right] dr \, .
 \label{step-intermedio}
\eeq
The  left hand side in eq.~(\ref{step-intermedio})
is  zero, because the surface term at the horizon
vanishes due to the $g$ factor and
the surface term at $\infty$ vanishes if $a'$ vanishes faster than $1/r$
(which is true if $\tilde{a}=0$). We thus find
\beq
\int_{r_h}^\infty \left[ {a'}^2  \le \frac{g \,  (1+r^2)}{r}  \ri  
 +2 e^2 \frac{(a-n)^2}{r} \, h \, H^2 \right] dr =0 \, .
\eeq
The function $h$ does not change sign for $r>0$, otherwise from the
third equation in the system in eq.~(\ref{eom-generiche-2}) 
a singularity in $H'$ or $a'$ would be needed.
As a consequence, $a=n$ outside the horizon.
 By analytic continuation,   $a$ vanishes also
inside the horizon.

\section{Soliton in the large $e$ limit }
\label{appe-section-large-e}

\begin{figure}[H]
\begin{center}
\includegraphics[scale=0.5]{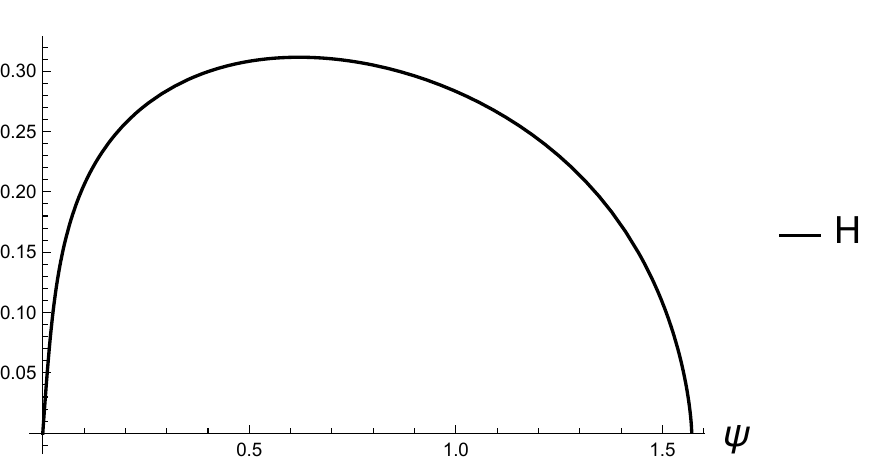} 
\qquad
\includegraphics[scale=0.5]{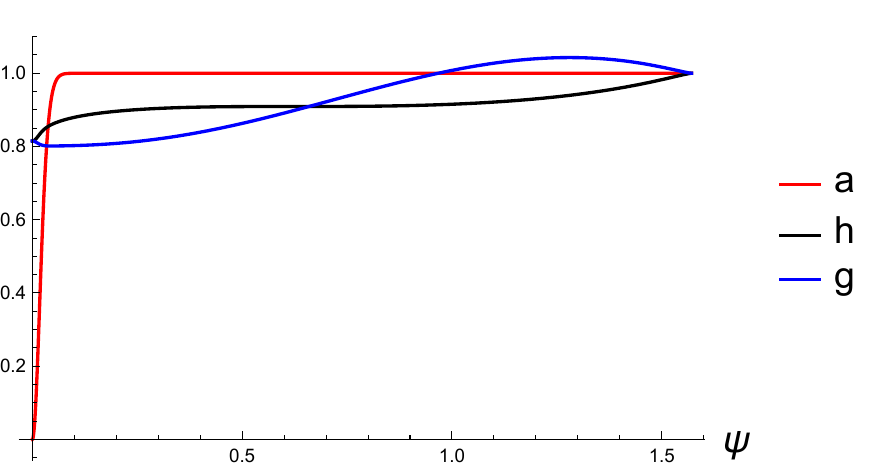} 
\caption{  \small  Example of soliton solution with $n=1$ and large $e$.
Here we set $e=500$, $m^2=-0.9$,  $G=0.1$, $\kappa=-1.000$.
 }
\label{esempio-vortex-large-e}
\end{center}
\end{figure}

In the large $e$ limit with fixed $\kappa$, $G$ and $n$,
the profile function $a$ tends to become equal to $n$
with the exception of a small region nearby the origin, where it vanishes.
In this case, the magnetic field is concentrated in a bulk
region which is much smaller compared to the AdS radius,
and so we expect that the behavior is similar to a vortex 
in flat space, for which a black hole horizon never forms.
Let us set for simplicity $n=1$.
In this case we expect that $a(r)=1$, with the exception of a very small 
region nearby the origin $r=0$, see figure~\ref{esempio-vortex-large-e}
for an example of solution.  

From eq.~(\ref{H-origine}), 
we have  $H(r) \approx H_0 \, r$ nearby $r=0$.
 We expect that $H_0$ gets large for $e \to \infty$.
 Nearby $r=0$, we can approximate the equation for $a$
 in eq.~(\ref{eom-generiche-2}) as follows
 \beq
 a''=\frac{a'}{r} +2 e^2 (a-1) \, H_0^2 r^2 \, .
 \eeq
 Imposing that $a$ vanishes at the origin and tends to $1$
 outside the central region, we find the solution
  \beq
 a  =1-\exp \le -\frac{H_0 e \, r^2}{\sqrt{2}} \ri \, .
 \eeq
The contribution to the energy density $\rho$ at the center
of the vortex scales as
\beq
\rho \approx 2 H_0^2 \, \exp(-\sqrt{2} \, H_0 e \, r^2) \, ,
\label{buca}
\eeq
which has a rectangular shape,
with height $\propto H_0^2$ and size $\propto \frac{1}{\sqrt{e H_0}}$.
If  $H_0$ would scale as $e$ at large  $e$, this would give a representation of
a $\delta$ function.  If  instead $H_0$ scales slower than $e$,
eq.~(\ref{buca}) tends to zero.
By numerical analysis for $G=0.1$ and $\kappa=-1$, we find 
$H_0 \propto e^{0.7}$.
This suggests that, if we fix $\kappa$ and increase $e$
we do not have a critical $e$ for which 
the soliton solutions becomes a black hole.
This is confirmed by the numerical solution
in figure~\ref{esempio-vortex-large-e}.

\end{document}